\documentclass[12pt,letterpaper]{article} 

\usepackage[english]{babel} 
\usepackage[utf8x]{inputenc} 

\usepackage{lipsum} 
\usepackage{setspace} 
\usepackage{microtype} 

\usepackage[margin=1.5in]{geometry} 
\usepackage{changepage} 

\usepackage{amsmath} 
\usepackage{amssymb} 
\usepackage{bm} 
\usepackage[retainorgcmds]{IEEEtrantools} 

\usepackage{graphicx} 
\usepackage[font=normalsize]{caption} 
\usepackage{subcaption} 

\usepackage{booktabs} 
\usepackage{tabularx} 
\usepackage{longtable} 
\usepackage{adjustbox} 
\usepackage{tabularray} 
\usepackage{multirow} 

\usepackage{listings} 
\usepackage{verbatim} 

\usepackage[colorlinks=true]{hyperref} 
\hypersetup{ 
  colorlinks,
  citecolor=blue,
  linkcolor=red,
  urlcolor=cyan
  }

\usepackage[colorinlistoftodos]{todonotes} 
\usepackage{soul} 
\sethlcolor{yellow} 

\usepackage{enumitem} 
\usepackage{pifont} 

\usepackage{xcolor} 
\usepackage{float} 
\usepackage{ragged2e} 
\usepackage{lineno} 
\usepackage{multicol} 
\usepackage{pdflscape} 
\usepackage{lscape} 
\usepackage[super]{nth} 
\usepackage{eurosym} 

\usepackage{natbib} 

\usepackage{authblk}

\usepackage{fancyhdr} 
\fancyhfoffset[L]{0cm} 
\fancyhfoffset[R]{0cm} 
\fancypagestyle{abstractstyle}{
    \fancyhf{} 
    \fancyfoot[C]{\textcolor{gray}{\thepage}}

}

\begin{document}


\vspace{-5cm}

\title{
\textbf{The Italian Municipality Equitable and Sustainable Well-being Index (MESWI)}
}

\newcommand{\thetitle}{MESWI}


\author{\href{https://orcid.org/0000-0002-9621-4008}{\includegraphics[width=0.35cm]{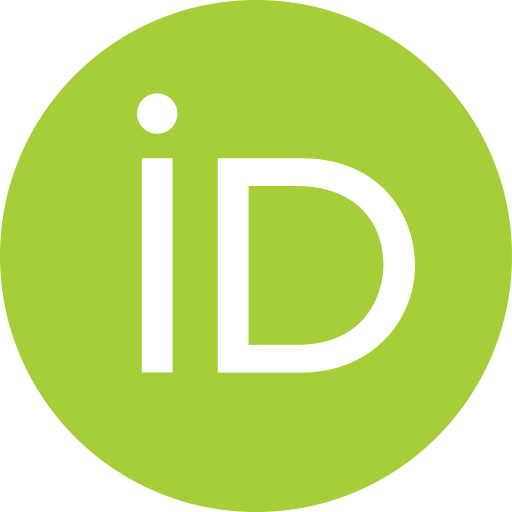}} Nicola Caravaggio$^{a,\dag,*}$} 
\author{\href{https://orcid.org/0000-0002-3913-0510}{\includegraphics[width=0.35cm]{Figures/orcid_id.png}} Giuliano Resce$^{a}$} 
\author{\href{https://orcid.org/0009-0004-5369-1594}{\includegraphics[width=0.35cm]{Figures/orcid_id.png}} Agapito Emanuele Santangelo$^{a}$} 


\affil{\footnotesize
  $^{a}$\textit{Department of Economics, University of Molise, Campobasso -- Italy} 
  \\
  \vspace{0.3cm}
  \begin{minipage}{11cm}~\\
  \centering
  \footnotesize
  $^{\dag}$\tt\href{mailto:nicola.caravaggio@unimol.it}{nicola.caravaggio@unimol.it} 
  \vspace{-1cm} 
  \end{minipage}
  }

\date{\small\today}

\renewcommand\Authands{, and } 

\maketitle


\begin{adjustwidth}{-0.1in}{-0.1in} 
\begin{abstract}

\thispagestyle{abstractstyle}

\noindent\rule[0.5ex]{\linewidth}{0.5pt}

\footnotesize{


This paper develops the Municipal Equitable and Sustainable Well-being Index (MESWI) for all Italian municipalities, extending the 12-domain BES framework to the local level through 49 indicators. Results reveal substantial territorial heterogeneity that regional and provincial statistics may conceal. The North-South divide remains the dominant geographical pattern and is considerably stronger than the difference between inner and non-inner areas. At the same time, municipalities within the same regions and territorial categories display markedly different well-being profiles. Remoteness is associated with weaker services and socio-economic opportunities but also with stronger environmental performance. The MESWI provides a fine-grained picture of multidimensional well-being and a potential information base for the programming, monitoring and evaluation of place-based policies.

}

\noindent\rule[0.5ex]{\linewidth}{0.5pt}

\end{abstract}
\end{adjustwidth}

\begin{quote}

	

    
 
\end{quote}

\pagebreak


\onehalfspacing 


\section{Introduction}
\label{section:i} 

Economic output remains indispensable for assessing productive capacity, yet it provides only a partial account of how people live. The limitations of gross domestic product (GDP) as a measure of social progress have been recognised since the earliest debates on national accounting \citep{Kuznets1934} and became particularly prominent with the empirical literature on income and happiness \citep{Easterlin1974}. The capability approach subsequently shifted attention from the possession of resources to the substantive freedoms that individuals are able to exercise \citep{Sen1985,Sen1999}. This broader perspective was consolidated by the Commission on the Measurement of Economic Performance and Social Progress (CMEPSP), which called for the joint consideration of material living conditions, health, education, environmental quality, social relations, security, and subjective evaluations \citep{Stiglitz2009}.

Moving beyond GDP has an important implication for territorial analysis: well-being must be understood as intrinsically multidimensional. A locality may perform favourably in terms of income or employment while exhibiting serious deficiencies in access to services, environmental conditions, or institutional capacity. Conversely, economically fragile territories may possess environmental, cultural, or social assets that are not captured by conventional measures of economic performance \citep{greco2020measuring}. Growing recognition of these limitations has prompted international institutions to develop a range of multidimensional frameworks aimed at capturing aspects of well-being that conventional economic indicators fail to reflect \citep{costanza2016modelling}. Among the most prominent examples are the Human Development Index (HDI), introduced by the United Nations Development Programme (\citeauthor{undp1990human}) in \citeyear{undp1990human}, and the OECD (Organisation for Economic Co-operation and Development) Better Life Index (BLI), launched in 2011 \citep{durand2015oecd}. These initiatives reflect a broader shift from measuring economic performance alone towards assessing well-being through multiple dimensions of individual and societal conditions.

Alongside these international initiatives, several countries have developed national well-being frameworks that adapt the beyond-GDP agenda to their specific institutional and statistical contexts. Prominent examples include the UK Measures of National Well-being, started in 2011, \citep{ONS2026}, the Dutch Monitor of Well-being, in 2018 \citep{CBS2026}, and the Quality of Life Framework for Canada, in 2021, \citep{StatisticsCanada2023}. Italy has followed a similar path through the development of the Equitable and Sustainable Well-being framework (\textit{Benessere Equo e Sostenibile} -- BES). Developed jointly by the Italian National Institute of Statistics (ISTAT) and the National Council for Economics and Labour (\textit{Consiglio Nazionale dell'Economia e del Lavoro} -- CNEL), the BES represents one of the most advanced institutional attempts to incorporate multidimensional well-being into official statistics and public decision-making \citep{ISTATCNEL2013,ISTAT2025b}. The framework organises quality of life into 12 domains and has stimulated a substantial methodological literature on indicator selection, normalisation, weighting and aggregation \citep{MazziottaPareto2018,Facchinetti2022}. Nevertheless, most official BES indicators remain available at the national, regional (NUTS 2) or provincial (NUTS 3) level. A gap therefore persists between the conceptual breadth of the framework and the territorial granularity required to diagnose local conditions and design place-based policies \citep{barca_2009}.

This gap is particularly consequential for the design and implementation of targeted policies, which increasingly operate at fine territorial scales. This is especially relevant in Italy, where municipalities range from large metropolitan centres to very small mountain and island communities, and such heterogeneity is only imperfectly captured by statistics produced at higher levels of territorial aggregation. Italian municipalities are key actors in the production and delivery of goods and services connected to growth and sustainable development. Italian municipalities have historically suffered from evident heterogeneity in terms of administrative
capacity and bureaucratic performance, partly due to the unresolved dualism between northern and southern regions, partly due to the presence of large fiscal imbalances among the different orders of government and territories \citep{patrizii2015public,antulov2021predicting,resce2022impact,caravaggio2026predicting}. Italy's territorial heterogeneity extends well beyond the traditional North-South divide, encompassing pronounced differences between urban and rural areas, inner areas and service centres, as well as municipalities of vastly different population sizes \citep{monturano2025short,bergantino2026far}. These differences translate into highly heterogeneous local economic and social conditions, access to services, administrative capacity and policy needs. Municipalities therefore represent a necessary scale of analysis for identifying territorial disparities that regional or provincial aggregates may conceal and for effectively targeting place-based policies.

Against this background, this paper develops a multidimensional measure of equitable and sustainable well-being for the entire universe of Italian municipalities, providing a local-scale counterpart to the BES framework.

The relevance of such fine-grained measurement is particularly evident in the case of Italy's inner areas. The National Strategy for Inner Areas (\textit{Strategia Nazionale per le Aree Interne} -- SNAI) classifies municipalities primarily according to travel time from service centres providing essential education, healthcare, and transport functions \citep{uval2014strategia,lucatelli2022}. The strategy combines interventions aimed at improving essential services with locally designed development projects and has been maintained across the 2014--2020 and 2021--2027 cohesion-policy cycles. It is complemented by other place-based programmes, including investments under the National Recovery and Resilience Plan (\textit{Piano Nazionale di Ripresa e Resilienza} -- PNRR) for the cultural and economic regeneration of small villages \citep{ItaliaDomani2026borghi}. These policies recognise that demographic decline, labour-market weakness and service deprivation are spatially concentrated and therefore require territorially differentiated responses and measures at the municipal level. 

A growing literature has moved well-being measurement towards finer spatial scales. Provincial applications of BES have shown that territorial rankings can depend materially on the aggregation procedure adopted \citep{Ciommi2017}. Urban studies have combined the BES conceptual structure with efficiency and entropy-based methods \citep{Nissi2018}. More recent contributions demonstrate the feasibility of municipality-level composite measures of well-being, competitiveness, administrative quality, or rurality \citep{Mazziotta2019,Fioroni2021,Scaccabarozzi2024,aiello2025geografia,cerqua2025municipal}. These studies provide important advances, but they generally employ a restricted set of well-being dimensions, focus on a more specific territorial construct, or are not designed to reproduce the complete BES architecture across the full municipal universe.

This paper addresses this gap by constructing the Municipal Equitable and Sustainable Well-being Index (MESWI) for all 7,896 Italian municipalities (as of December 31, 2024). Its contribution is threefold. First, it operationalises the 12-domain BES framework at the local administrative-unit level using 49 indicators obtained from statistical and administrative sources. Second, it employs a transparent hierarchical aggregation procedure based on the Adjusted Mazziotta--Pareto Index, preserving the conceptual role of the domains while penalising markedly unbalanced well-being profiles. Third, it uses the resulting index to examine the relationship between municipal well-being, inner-area status and Italy's broader macro-regional divide.

The analysis is organised around three questions: (i) how much does municipal well-being vary across the North, Centre, and South and Islands?; (ii) How large is the additional difference associated with inner-area status?; (iii) Which domains characterise the well-being profiles of increasingly remote municipalities? 

The results reveal a persistent divide between the South and the rest of the country, substantial heterogeneity within every region, and a statistically detectable but substantively smaller difference associated with inner-area status. They also show that remoteness is linked to weaker performance in services, innovation, and several socio-economic domains, while peripheral municipalities display a systematic environmental advantage. The latter should be interpreted as a measurable territorial asset, although its capacity to support employment and demographic resilience depends on local institutions, productive linkages and cooperation. Overall, the findings support the use of multidimensional municipal indicators as a complement---rather than a substitute---to accessibility-based territorial classifications.

The remainder of the paper is organised as follows. Section~\ref{section:ii} describes the institutional setting of Italy's Inner areas. Section~\ref{section:iii} presents the data and the construction of the index. Section~\ref{section:iv} reports and discusses the empirical results and their policy implications. Section~\ref{section:v} concludes.


\section{Territorial segmentation and Italy's inner areas}
\label{section:ii}

Italian territorial disparities are characterised by multiple and partly overlapping geographical divisions. The most established is the North-South divide, which reflects persistent differences in economic performance, labour-market opportunities, public-service provision, and institutional capacity \citep{greco2018stochastic,antulov2021predicting,mauro2023decentralization, cerqua2025municipal,di2025determinants,monturano2025income}. Yet this macro-regional cleavage does not exhaust the geography of territorial disadvantage. Substantial heterogeneity also exists within regions, reflecting differences in settlement patterns, population size, accessibility and proximity to essential services \citep{di2025regional,monturano2025short,bergantino2026far,mariani2026place}. In this respect, a second relevant dimension is the centre--periphery divide, which cuts across the traditional North-South geography and distinguishes municipalities according to their degree of integration into networks of services and economic opportunities. The two dimensions are related but conceptually distinct: a remote municipality in the North may face important accessibility constraints while maintaining relatively favourable socio-economic conditions, whereas a more accessible municipality in the South may still experience broad multidimensional disadvantage. Territorial segmentation in Italy therefore cannot be adequately represented by a single geographical classification.

The National Strategy for Inner Areas (\textit{Strategia Nazionale per le Aree Interne} -- SNAI) provides the main institutional expression of this centre--periphery dimension. Designed between 2012 and 2014, implemented during the 2014--2020 cohesion-policy cycle and subsequently retained for 2021--2027, the SNAI adopts an explicitly \textit{place-based} approach in which territorial policies are expected to respond to the specific economic, social, demographic, and institutional characteristics of individual places \citep{barca_2009,barca2012case,barca2019place}. Rather than identifying disadvantaged territories solely through administrative, demographic or altimetric criteria, the Strategy defines ``innerness'' primarily in relation to effective accessibility to essential services \citep{uval2014strategia,dps2012aree,nuvap2020strategia}. Its underlying rationale is that access to education, healthcare and mobility affects the effective exercise of citizenship rights and, more broadly, the ability of individuals to live, work and remain in a territory.

The resulting classification reflects the polycentric structure of the Italian settlement system. Municipalities providing the required education, healthcare, and transport services are classified as \textit{Hubs}, while neighbouring municipalities jointly supplying these functions may constitute \textit{Inter-municipal Hubs}. Other municipalities are classified according to road travel time to the nearest service centre as \textit{Belt}, \textit{Intermediate}, \textit{Peripheral}, or \textit{Ultra-peripheral}. Inner Areas conventionally comprise the latter three classes, whereas Poles, Inter-municipal Poles and Belt municipalities constitute the more accessible part of the country \citep{uval2014strategia}. Importantly, this makes ``innerness'' a relational measure of accessibility rather than a synonym for rurality, mountainous location or economic backwardness. Remoteness may coexist with demographic decline, weak labour demand or limited administrative capacity, but these characteristics do not formally determine SNAI status.

This distinction is important because peripheral territories cannot be understood exclusively in terms of deprivation. Alongside service scarcity, demographic contraction and fragile productive structures, many inner areas possess environmental, landscape, cultural, and productive assets that can provide a basis for locally differentiated development strategies \citep{uval2014strategia,bolzoni2026repopulating}. Consistently with this perspective, the SNAI combines two complementary lines of intervention: improving access to essential services and promoting local development through the mobilisation of natural and cultural resources, local production systems, tourism, skills, and community initiatives \citep{uval2014strategia,lucatelli2022,bolzoni2026repopulating}. Its implementation consequently relies on a multilevel governance model in which municipalities cooperate within territorial coalitions and formulate Area Strategies in coordination with regional and central administrations and local stakeholders \citep{uval2014strategia,cdd_snai_2023}. This inter-municipal dimension is particularly relevant in territories where small population size and administrative fragmentation can limit the technical and organisational capacity of individual municipalities.

The scale of the phenomenon makes this territorial segmentation particularly relevant. Under the classification adopted for the 2021--2027 programming period, 3{,}834 municipalities are classified as Inner Areas, accounting for 48.5\% of Italian municipalities and 58.8\% of the national territory, but only 22.7\% of the resident population \citep{nuvap2020strategia,istat_aree_interne,psnai_2025}. The classification of Inner Areas should nevertheless be distinguished from the identification of specific \textit{SNAI project areas}: while the former assigns municipalities to accessibility classes, the latter consist of selected coalitions of municipalities receiving dedicated policy support. During the first SNAI cycle (2014--2020), 72 target areas were identified, comprising a total of 1{,}094 municipalities. In the second cycle (2021--2027), the number of target areas increased to 123 (plus a dedicated project for small islands), covering 1{,}904 municipalities.

The broader policy framework has subsequently been reinforced by the National Strategic Plan for Inner Areas (\textit{Piano Strategico Nazionale delle Aree Interne} -- PSNAI), which seeks to integrate essential services, infrastructure, economic development, human capital and administrative capacity while placing territorial habitability and the ``right to remain'' at the centre of policy action \citep{psnai_2025}.

The centre--periphery geography identified by the SNAI, however, should not be interpreted as an alternative to the traditional North-South divide. Rather, the two segmentations capture different dimensions of Italian territorial inequality. Accessibility identifies a structural constraint associated with the location of municipalities relative to essential services, while local well-being also depends on economic, demographic, social, environmental and institutional conditions. Municipalities facing similar accessibility constraints may therefore experience markedly different outcomes, while municipalities outside the inner area perimeter may display comparable forms of deprivation. Local resources, governance quality and administrative capacity further contribute to this heterogeneity \citep{benassi2024local,compagnucci2024inner,monturano2025short,mariani2026place}. Moreover, territorial accessibility itself is not static: \cite{bergantino2026far}, recalculating distances from essential services five years after the classification developed by \cite{nuvap2020strategia}, document a 20\% increase in the number of municipalities classified as inner areas.

This coexistence of macro-regional and centre-periphery disparities provides a strong rationale for analysing well-being at the municipal level. A composite municipal indicator does not replace the SNAI accessibility classification, nor does it supersede the established North-South geography. Instead, it provides a complementary dimension through which these territorial segmentations can be jointly examined. Measuring multidimensional well-being at the municipal scale makes it possible to assess whether increasing ``peripherality'' is systematically associated with poorer achieved well-being, whether this relationship differs across the broader North-South divide, and how municipalities facing similar accessibility constraints differ in the composition of their local advantages and disadvantages. In this sense, municipal multidimensional measurement provides the level of territorial detail required to move from broad geographical categories towards a more differentiated diagnosis of local conditions and, ultimately, more precisely targeted place-based policies.

\section{Data and methodology}
\label{section:iii} 

\subsection{Data sources}
\label{section:iii_1}

The analysis covers 7,896 municipalities (as of December 31, 2024) and uses the most recent observation available for each variable, generally from 2022 to 2025. The main source is \citeauthor{istat_amdc}'s \emph{A Misura di Comune} database, which integrates administrative and experimental statistics at municipal scale \citep{istat_amdc}. Additional information comes from the regional BES reports \citep{ISTAT2025a}, the national BES report \citep{ISTAT2025b}, WorldCover from European Space Agency \citep{ESA2026}, the Municipal Administration Quality Index (MAQI) of \cite{cerqua2025municipal}, and municipal travel-time measures obtained from an \textit{ad hoc} procedure described in the Appendix \ref{subsection:appendix_1}.

The baseline index contains 49 elementary indicators assigned to the 12 BES domains: (1) health; (2) education and training; (3) work and work--life balance; (4) economic well-being; (5) social relationships; (6) politics and institutions; (7) safety; (8) subjective well-being; (9) landscape and cultural heritage; (10) environment; (11) innovation, research and creativity; (12) quality of services. Table~\ref{tab:architecture} summarises the architecture, while Table~\ref{tab:indicator-details} in the Appendix \ref{subsection:appendix_2} provides the complete indicator list, reference year, source, polarity, and descriptive statistics.

\begin{table}[H]
\centering
\caption{\textit{MESWI domain architecture and spatial granularity of the source data}}
\label{tab:architecture}
\small
\begin{tabularx}{\textwidth}{@{}l c X@{}}
\toprule
\multicolumn{1}{c}{\textbf{Domain}} &
\multicolumn{1}{c}{\textbf{Indicators}} &
\multicolumn{1}{c}{\textbf{Spatial granularity}} \\
\midrule
Health & 3 & Municipal \\
Education and training & 5 & Municipal; selected indicators imputed from provincial-capital values \\
Work and work-life balance & 4 & Municipal \\
Economic well-being & 4 & Municipal; selected indicators available above 5,000 residents \\
Social relationships & 3 & Municipal; selected indicators available above 5,000 residents \\
Politics and institutions & 5 & Municipal \\
Safety & 6 & Provincial BES \\
Subjective well-being & 4 & Regional BES \\
Landscape and cultural heritage & 3 & Municipal \\
Environment & 5 & Municipal; selected indicators imputed from provincial-capital values \\
Innovation, research and creativity & 2 & Municipal \\
Quality of services & 5 & Municipal \\
\bottomrule
\end{tabularx}
    \begin{minipage}{0.96\textwidth}
    \vspace{.1cm}
    \justifying
    \tiny
    \noindent
    Notes: the table summarises the 49 elementary indicators used in the baseline index. The complete list and descriptive statistics are reported in Table~\ref{tab:indicator-details}.
    \end{minipage}
\end{table}

The use of municipal data requires several transparent imputations. For variables released only for provincial capitals or municipalities above 5,000 residents, uncovered municipalities are assigned the corresponding provincial value. Safety indicators are observed at provincial level and subjective well-being at regional level, so every municipality within the same province or region receives the same value for those domains. Remaining sporadic missing observations are replaced by the indicator median. This strategy preserves nationwide coverage, but it also implies that some dimensions contain less within-region variation. We therefore treat the 12-domain index as the baseline and explicitly test sensitivity to removing safety and subjective well-being.

\subsection{Construction of the indicator}
\label{section:iii_2}


Composite indicators reduce a multidimensional information set to a synthetic measure, but their construction embeds substantive judgements. The choice of domains defines the phenomenon; polarity determines what counts as improvement; normalisation shapes relative distances; weighting encodes the importance of components; aggregation governs the degree of compensation between them \citep{OECD2008,MazziottaPareto2013}. A high value in one dimension can fully offset a poor value in another under an arithmetic mean, whereas non-compensatory approaches penalise unbalanced profiles\citep{ResceMaynard2018,greco2020measuring,Resce2021}. The latter feature is especially relevant for well-being. For example, a municipality with high cultural provision but severe service deprivation should not necessarily receive the same overall score as a municipality with a similar mean but a more balanced profile. The Adjusted Mazziotta--Pareto Index (AMPI) was developed to combine a central tendency with a penalty for horizontal imbalance across component indicators \citep{MazziottaPareto2016,MazziottaPareto2018}. It is widely used in Italian territorial statistics because it remains transparent, does not require statistically estimated weights, and limits complete substitution between dimensions (\textit{e.g.}, \citealp{Ciommi2017,Fioroni2021,Scaccabarozzi2024,cerqua2025municipal}).

Local measurement also faces a data problem. Municipal indicators are increasingly available through administrative sources, but coverage remains uneven across domains. Some variables are observed only for municipalities above a population threshold or for provincial capitals; others are released at provincial or regional level. Constructing a national municipal index therefore requires explicit decisions about spatial imputation and sensitivity analysis. The MESWI addresses these constraints through a baseline 12-domain specification and a robustness version that excludes the two domains with exclusively supra-municipal information. Our indicator is constructed hierarchically: elementary indicators are first harmonised and aggregated within each domain; the 12 domain scores are then combined through an AMPI-based non-compensatory rule.


Let $x_{ij}$ denote the value of indicator $j$ in municipality $i$ where $j$ spans from 1 to 49, the number of municipality variables selected. Indicators are oriented so that higher values always represent higher well-being. Variables with negative polarity are multiplied by $-1$. Strongly right-skewed non-negative indicators are transformed as follows:
\begin{equation}
 x'_{ij}=\log(1+x_{ij})
 \label{eq:log}
\end{equation}
which preserves ordering while compressing extreme ratios. This transformation is applied to the numbers of libraries, cultural institutions and museum visitors per 100,000 residents, productive specialisation in high-technology sectors, and children enrolled in municipal childcare services.\footnote{The indicator is calculated as the number of children aged 0--2 who used municipal childcare services, including nurseries, micro-nurseries, and supplementary early-childhood services, divided by the average annual resident population aged 0--2, multiplied by 100.}

Each oriented indicator is then converted to a $z$-score:
\begin{equation}
 z_{ij}=\frac{x'_{ij}-\bar{x}_{j}}{s_j}
 \label{eq:zscore}
\end{equation}
where $\bar{x}_j$ and $s_j$ are the cross-municipality mean and standard deviation, respectively. Standardisation removes the original unit of measurement and places all elementary indicators on a comparable scale.


For domain $k$, containing $m_k$ indicators, the municipal domain score is the following arithmetic mean:
\begin{equation}
 D_{ik}=\frac{1}{m_k}\sum_{j\in k} z_{ij}
 \label{eq:domain}
\end{equation}
Equal weights are used within domains for transparency and to avoid assigning data-driven importance that may conflict with the normative structure of BES indicators. The hierarchical design also prevents domains with more elementary indicators from automatically receiving greater weight in the final index.


The domain distributions contain structurally extreme municipalities, especially for per-capita cultural facilities and visitors. Using the observed minimum and maximum as normalisation anchors would allow a small number of outliers to compress most municipalities into a narrow interval. We therefore define empirical goalposts as the 5\textsuperscript{th} and 95\textsuperscript{th} percentiles of each domain distribution. Let $g_{5,k}$ and $g_{95,k}$ denote these thresholds. The domain score is first \textit{winsorised}:
\begin{equation}
 \widetilde{D}_{ik}=\min\left\{g_{95,k},\max\left(g_{5,k},D_{ik}\right)\right\}
 \label{eq:winsor}
\end{equation}
and then mapped to the conventional 70--130 AMPI scale:
\begin{equation}
 D^{*}_{ik}=70+60\frac{\widetilde{D}_{ik}-g_{5,k}}{g_{95,k}-g_{5,k}}
 \label{eq:goalpost}
\end{equation}
The percentile choice retains 90\% of the observed domain range while bounding the leverage of exceptional observations. It is a robust empirical adaptation of the goalpost logic used in AMPI normalisation. Because the anchors are estimated from the present cross-section, future inter-temporal applications should keep them fixed over time.

For municipality $i$, let $K=12$ denote the number of well-being domains. The arithmetic mean and the within-municipality standard deviation of the normalised domain scores are defined as follows:\\
\begin{minipage}{0.48\textwidth}
\begin{equation}
M_i = \frac{1}{K}\sum_{k=1}^{K} D^{*}_{ik}
\label{eq:domain-mean}
\end{equation}
\end{minipage}
\hfill
\begin{minipage}{0.48\textwidth}
\begin{equation}
S_i =
\sqrt{
\frac{1}{K-1}
\sum_{k=1}^{K}
\left(D^{*}_{ik}-M_i\right)^2
}
\label{eq:domain-sd}
\end{equation}
\end{minipage}\\
The relative dispersion of the municipality's domain profile is measured by the following coefficient of variation:
\begin{equation}
CV_i=\frac{S_i}{M_i}
\label{eq:domain-cv}
\end{equation}
Finally, the positive version of the composite index is then computed as follows: 
\begin{equation}
MESWI_i  = M_i-S_iCV_i = M_i-\frac{S_i^2}{M_i}
\label{eq:besc}
\end{equation}
where higher values indicate greater multidimensional well-being. The second
term represents a penalty for imbalance across domains: for a given average
level of well-being, the penalty increases with the dispersion of the domain
scores. Consequently, two municipalities with the same mean performance may
receive different index values when one displays substantially greater
shortfalls in particular domains. The aggregation procedure is therefore
partially non-compensatory, since outstanding performance in a limited number
of domains cannot fully offset weak performance elsewhere.

\subsection{Inner-area classification and empirical comparisons}
\label{section:iii_3}

The empirical analysis combines the MESWI with the 2021--2027 SNAI classification \citep{nuvap2020strategia}. A municipality is classified as an inner-area municipality when it belongs to one of the intermediate, peripheral, or ultra-peripheral accessibility classes, or when it is included in an officially designated SNAI project area.\footnote{It should be noted that 209 municipalities classified as belt areas are also included in this category. In addition, one hub municipality, Orvieto, falls within a SNAI-designated inner area.} This definition is used throughout the analysis to ensure consistency between the descriptive comparisons and the subsequent statistical tests.

The purpose of this classification is not to redefine the SNAI geography, but to assess whether municipalities characterised by weaker accessibility to essential services also exhibit systematically different levels of multidimensional well-being. In this respect, the SNAI classification and the MESWI capture two related but distinct territorial dimensions. The former is primarily based on accessibility to education, healthcare, and transport services, whereas the latter summarises a broader set of economic, social, institutional, environmental, and service-related conditions. Their joint analysis therefore makes it possible to evaluate whether territorial remoteness is associated with a wider well-being disadvantage and whether such an association differs across the country.

For the national descriptive comparison, municipalities are divided into inner areas and other municipalities. For the domain-profile analysis, however, all six official SNAI accessibility classes are retained: hub, inter-municipal hub, belt, intermediate, peripheral, and ultra-peripheral. This more detailed classification is used to investigate whether the individual well-being domains display a gradual pattern as accessibility decreases, rather than a simple dichotomy between inner and non-Inner areas.

Municipalities are also grouped into three broad macro-regions: North, Centre, and South and Islands (\textit{Mezzogiorno}). This second territorial dimension is included because the descriptive evidence points to substantial and historically persistent geographical disparities in well-being across Italy \citep{mauro2023decentralization}. The empirical strategy is therefore designed to distinguish the association between inner-area status and municipal well-being from the broader North-South divide.

To formalise these comparisons, we estimate a two-way analysis of variance (ANOVA) model:
\begin{equation}
MESWI_i
=
\mu
+
\alpha_{r(i)}
+
\beta_{s(i)}
+
(\alpha\beta)_{r(i),s(i)}
+
\varepsilon_i
\label{eq:anova}
\end{equation}
where $MESWI_i$ denotes the well-being score of municipality $i$, $\mu$ is the overall mean, $\alpha_{r(i)}$ captures the effect associated with the macro-region of municipality $i$, and $\beta_{s(i)}$ captures the effect associated with inner-area status. The interaction term $(\alpha\beta)_{r(i),s(i)}$ allows the inner-area differential to vary across macro-regions, while $\varepsilon_i$ represents the residual component not accounted for by the observed grouping structure.

The model is used for two related purposes. First, it tests whether the mean differences observed in the descriptive figures are statistically distinguishable across macro-regions and between inner and non-inner municipalities. Second, it decomposes the total variability of the index into the components associated with macro-regional location, inner-area status, their interaction, and the residual within-group variation:
\begin{equation}
SS_{\mathrm{Total}}
=
SS_{\mathrm{Macroregion}}
+
SS_{\mathrm{InnerArea}}
+
SS_{\mathrm{Interaction}}
+
SS_{\mathrm{Residual}}
\label{eq:ss-decomposition}
\end{equation}
For each model component, the corresponding $F$ statistic is calculated as the ratio between the relevant between-group mean square and the residual mean square:
\begin{equation}
F
=
\frac{MS_{\mathrm{Between}}}
     {MS_{\mathrm{Within}}}
\label{eq:f-statistic}
\end{equation}
where each mean square is obtained by dividing the corresponding sum of squares by its degrees of freedom.\footnote{The analysis was conducted entirely in \href{https://www.r-project.org/}{R}.}


\section{Results}
\label{section:iv} 

\begin{figure}[H]
\centering
\caption{\textit{Distribution of MESWI by macro-region and inner-area status}}
\includegraphics[width=0.94\textwidth]{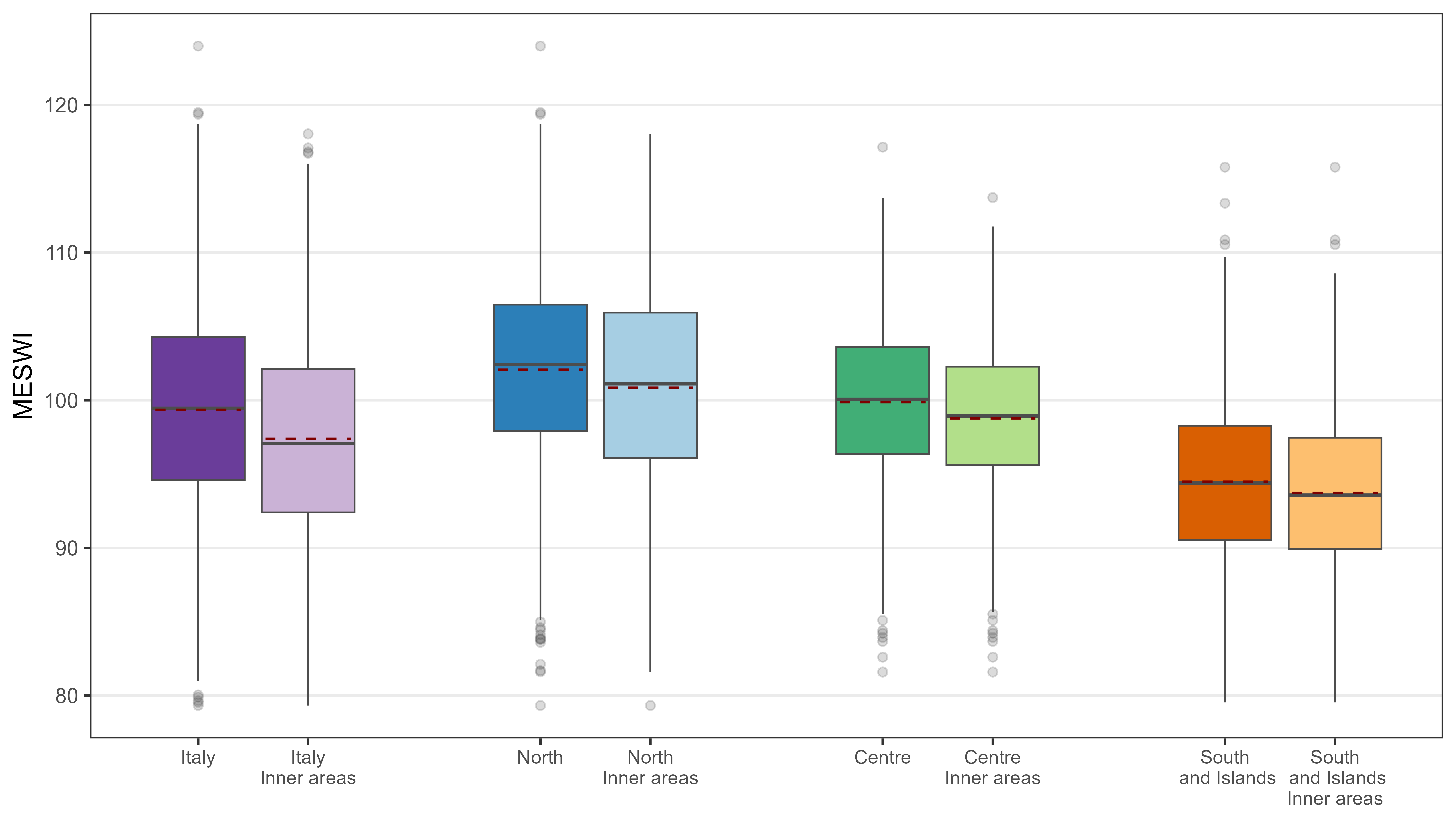}
\label{fig:boxplot}
\begin{minipage}{0.92\textwidth}
\tiny
Notes: inner areas comprise intermediate, peripheral and ultra-peripheral municipalities, together with municipalities belonging to SNAI project areas under the 2021--2027 classification. The solid grey line inside each box represents the median, while the dashed dark-red line indicates the mean. Boxes report the interquartile range, whiskers extend to 1.5 times the interquartile range, and points denote outlying observations.
\end{minipage}
\end{figure}

Figure~\ref{fig:boxplot} reports the distribution of MESWI values for Italy as a whole, for the three macro-regions (North, Centre, and South and Islands), and for the corresponding inner-area subsets. The box-plots confirm a clear territorial gradient. Municipalities in the North display the highest central values, followed by those in the Centre, whereas the South and Islands is characterised by a markedly lower distribution. This pattern is visible not only in the means, shown by the dashed red lines, but also in the medians and interquartile ranges.

Within each macro-region, inner-area municipalities generally exhibit lower MESWI values than the corresponding overall distribution. However, these within-region differences are relatively modest compared with the broader geographical divide, particularly that separating the South and Islands from the Centre and the North. The substantial overlap between the distributions also indicates that inner-area status does not identify a homogeneous group of low-well-being municipalities. In each macro-region, some inner-area municipalities attain MESWI values comparable to those of non-inner municipalities, while some municipalities outside Inner areas perform relatively poorly.

Figure~\ref{fig:boxplot} also highlights considerable within-group heterogeneity, especially in the North and Centre, where both low- and high-value outliers are present. By contrast, the distribution for the South and Islands is shifted downward and appears somewhat more compressed. Overall, the box-plots suggest that macro-regional location accounts for a larger share of territorial differences in municipal well-being than inner-area status alone, while also revealing substantial variation among municipalities within the same territorial category.

The two-way ANOVA (ANanlysis Of VAriance) in Table~\ref{tab:anova-main} confirms this ordering but also clarifies its relative importance \citep{Fisher1925,Montgomery2020}. Macro-regional location explains 27\% of the residual-adjusted variation in MESWI, compared with 3\% for inner-area status. The interaction is statistically significant at the 5\% level but substantively negligible \citep{Lakens2013}. Thus, the mean disadvantage associated with inner-area status is broadly similar across the country, while the much larger divide is between the South and the rest of Italy.

\begin{table}[htbp]
\centering
\caption{\textit{Two-way analysis of MESWI's variance}}
\label{tab:anova-main}
\small
\resizebox{\textwidth}{!}{%
\begin{tabular}{@{}lrrrrrr@{}}
\toprule
\multicolumn{1}{c}{\textbf{Source of variation}} &
\multicolumn{1}{c}{\textbf{df}} &
\multicolumn{1}{c}{\textbf{Sum of squares}} &
\multicolumn{1}{c}{\textbf{Mean square}} &
\multicolumn{1}{c}{\textbf{$F$}} &
\multicolumn{1}{c}{\textbf{$p$ value}} &
\multicolumn{1}{c}{\textbf{Partial $\eta^2$}} \\
\midrule
Macro-region & 2 & 92,894 & 46,447 & 1,435.15 & $<0.001$ & 0.270 \\
Inner-area status & 1 & 9,088 & 9,088 & 280.81 & $<0.001$ & 0.030 \\
Macro-region $\times$ inner-area status & 2 & 207 & 104 & 3.20 & 0.041 & 0.001 \\
Residuals & 7,893 & 255,448 & 32 & -- & -- & -- \\
\bottomrule
\end{tabular}%
}
\vspace{0.4em}
\begin{minipage}{\textwidth}
\vspace{.1cm}
\tiny
Notes: sequential sums of squares from the fitted factorial model. Effect sizes are partial $\eta^2$.
\end{minipage}
\end{table}

Furthermore, Tukey-adjusted comparisons \citep{Tukey1949} show that all three
macro-regional means differ in the baseline specification (Table~\ref{tab:tukey-main}). The Centre is 2.18 points below the North, the South and Islands is 7.58 points below the North, and it is 5.39 points below the Centre. The confidence intervals are narrow because the analysis covers the municipal universe.

\begin{table}[htbp]
\centering
\caption{\textit{Tukey-adjusted comparisons between macro-regions}}
\label{tab:tukey-main}
\small
\resizebox{\textwidth}{!}{%
\begin{tabular}{@{}lrrr@{}}
\toprule
\multicolumn{1}{c}{\textbf{Comparison}} &
\multicolumn{1}{c}{\textbf{Mean difference}} &
\multicolumn{1}{c}{\textbf{95\% confidence interval}} &
\multicolumn{1}{c}{\textbf{Adjusted $p$ value}} \\
\midrule
Centre $-$ North & -2.18 & $[-2.66,\,-1.71]$ & $<0.001$ \\
South and Islands $-$ North & -7.58 & $[-7.91,\,-7.25]$ & $<0.001$ \\
South and Islands $-$ Centre & -5.39 & $[-5.90,\,-4.89]$ & $<0.001$ \\
\bottomrule
\end{tabular}%
}
\vspace{0.4em}
\begin{minipage}{\textwidth}
\vspace{.1cm}
\tiny
Notes: mean differences are computed as the first macro-region minus the second. Adjusted $p$ values are based on \citeauthor{Tukey1949}'s procedure for multiple comparisons.
\end{minipage}
\end{table}

\subsection{Domain profiles along the SNAI accessibility gradient}
\label{section:iv_1}

\begin{figure}[H]
\centering
\caption{\textit{Mean standardised MESWI domain scores by SNAI class}}
\includegraphics[width=0.74\textwidth]{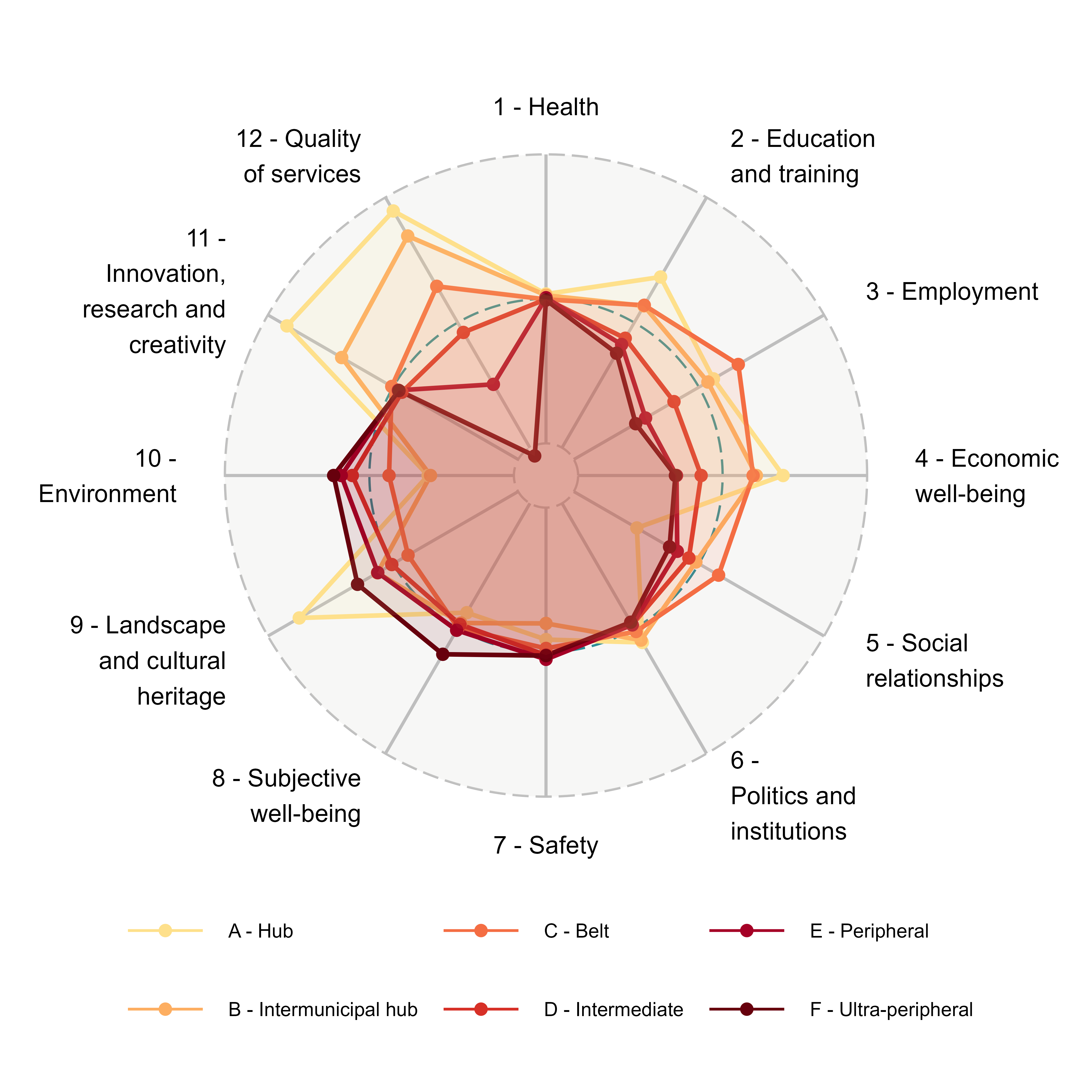}
\label{fig:radar}
\begin{minipage}{0.7\textwidth}
\tiny
Notes: dashed circle denotes the municipal mean (zero). Positive values indicate performance above the municipal mean after indicator polarity has been harmonised.
\end{minipage}
\end{figure}

Figure~\ref{fig:radar} compares mean standardised domain scores across the six SNAI classes through a radar graph. The largest and most systematic gradient appears in quality of services. The mean moves from 0.89 standard deviations above the municipal average in hubs to 1.07 standard deviations below it in ultra-peripheral municipalities. This result is partly mechanical, because the SNAI classification is itself based on access to schools, hospitals and transport. Its magnitude nevertheless confirms that the service-access indicators included in MESWI reproduce the intended territorial hierarchy.

A broader accessibility gradient is visible in education, economic well-being, and innovation. The latter is especially concentrated in hubs and inter-municipal hubs, reflecting stronger digital infrastructure and high-technology specialisation. Labour-market outcomes are less perfectly monotonic: belt municipalities have the highest mean score, while intermediate, peripheral and ultra-peripheral areas are clearly below average. This pattern is consistent with labour-market advantages in commuter belts around urban centres.

Other domains reveal trade-offs rather than a uniform hierarchy. Environmental quality improves as remoteness increases: hubs and inter-municipal hubs are below average, whereas intermediate and peripheral municipalities are above average. Landscape and cultural heritage displays a non-linear profile. Hubs benefit from dense cultural supply and visitor flows, while ultra-peripheral areas also perform relatively well because small populations coexist with substantial heritage endowments. Social relationships likewise do not follow a simple centre-periphery gradient: hubs score below average, belts above average, and the most remote classes moderately below average. Health varies little across classes. Safety and subjective well-being also exhibit limited or irregular differentiation, a result that must be interpreted cautiously because their source data are available only at provincial and regional level.

The radar therefore supports two conclusions. First, SNAI remoteness is associated with a multidimensional disadvantage that extends beyond the classification's service-access variables. Second, Inner areas are not uniformly worse off: they possess environmental and, in some cases, cultural advantages that are obscured by a single deficit-oriented narrative.

\subsection{National geography and regional heterogeneity}
\label{section:iv_2}

\begin{figure}[H]
\centering
\caption{\textit{MESWI deciles and SNAI project areas}}
\includegraphics[width=0.99\textwidth]{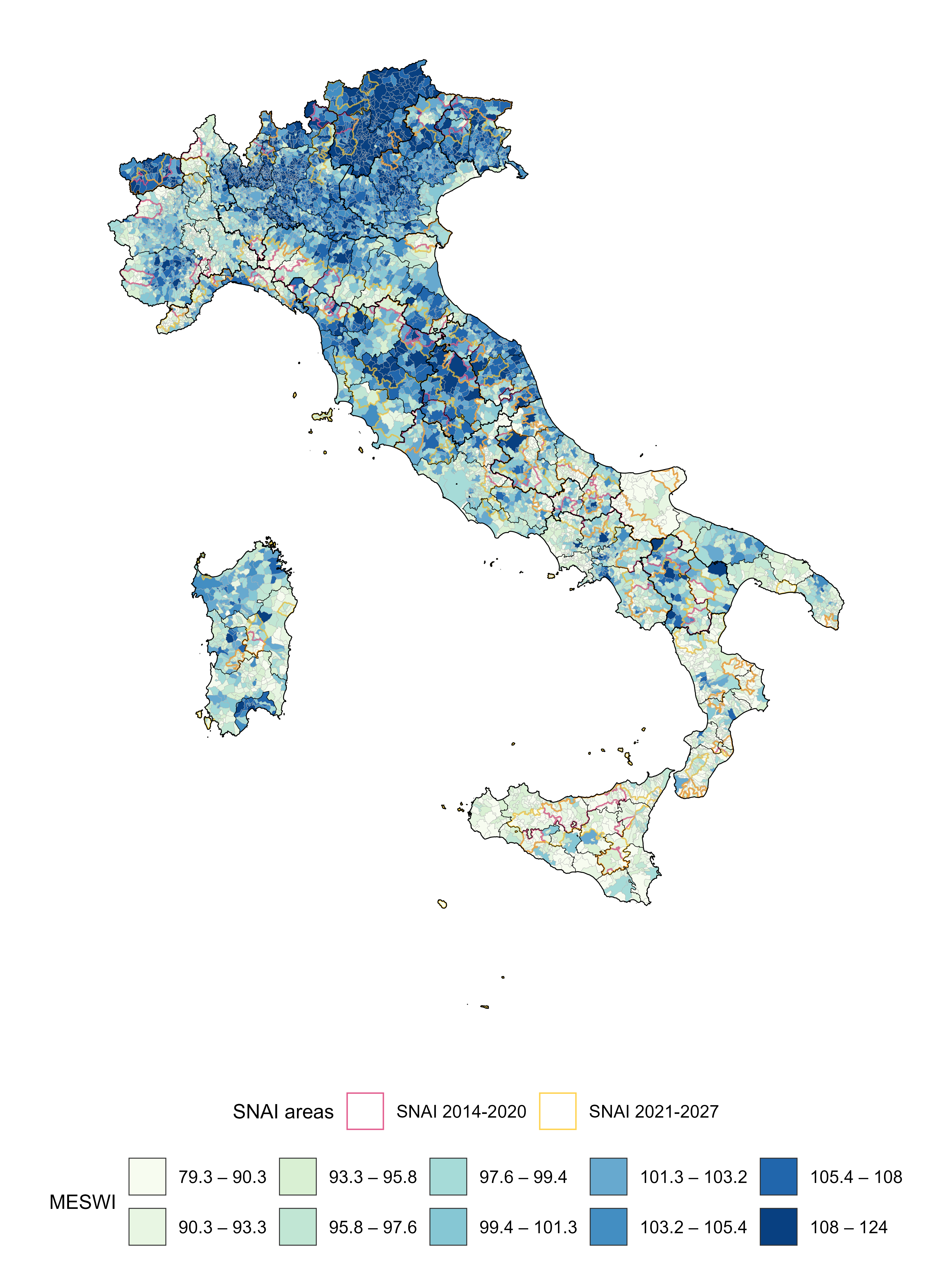}
\label{fig:national-map}
\begin{minipage}{0.90\textwidth}
\tiny
Notes: municipal fill colours represent national MESWI deciles. Pink and yellow outlines identify SNAI project areas in the 2014--2020 and 2021--2027 cycles, respectively.\end{minipage}
\end{figure}

The municipal map in Figure~\ref{fig:national-map} classifies MESWI into national deciles and overlays SNAI project-area boundaries for the two programming cycles. The broad spatial pattern is unmistakable. High-scoring municipalities are concentrated in the Alpine arc, the North-East, Lombardy, and parts of central Italy. Low-scoring municipalities are more frequent in Sicily, Calabria, and Apulia, instead. Yet the map also shows substantial local heterogeneity. High and low deciles coexist within every macro-region, and the national divide is crossed by numerous local exceptions. Maps for each region are provided in Figure \ref{fig:regions-map} in the Appendix~\ref{subsection:appendix_2}.

Regional means reinforce the spatial divide but also illustrate the information lost at aggregate scale. Trentino-Alto Adige records the highest unweighted municipal mean (109.83), followed by Valle d'Aosta (106.73), Friuli-Venezia Giulia (104.58), Lombardy (104.52) and Veneto (103.59). Sicily (90.91), Calabria (91.65) and Apulia (93.02) occupy the bottom of the distribution. The gap between the highest- and lowest-scoring regions is therefore almost 19 points. Table~\ref{tab:regions} in the Appendix~\ref{subsection:appendix_2} reports the complete regional results and distinguishes project areas, broader Inner areas and non-inner municipalities under both SNAI cycles.

At national level, the 2021--2027 broad inner area category averages approximately 97.3, compared with 101.4 among non-inner municipalities; designated project areas average about 96.6. The magnitude of the gap varies considerably by region. In some southern regions, both inner and non-inner municipalities perform relatively poorly, generating a small within region difference despite low overall well-being. In several northern and central regions, inner area municipalities remain above the national mean even though they underperform the other municipalities in their own region. This is why inner area status should not be interpreted as a universal low-welfare label.

\subsection{Robustness to spatial data granularity}
\label{section:iv_3}

The baseline index includes safety (7) and subjective well-being (8) to retain the complete BES architecture. These are also the two domains with the coarsest spatial information: safety is observed at provincial level while subjective well-being at regional level. Assigning common values to all municipalities within the same province or region suppresses local variation and may mechanically amplify differences between macro-regions. We therefore reconstruct the entire index after excluding these two domains. All transformations, goalposts and AMPI aggregation steps are repeated for the remaining ten domains.

The robustness results are informative rather than merely mechanical. The national spatial pattern remains stable, as shown by the reduced-index map Figure~\ref{fig:robust-map} in the Appendix~\ref{subsection:appendix_2}. The South and Islands remains substantially below the rest of the country, and most municipalities remain in the same or an adjacent decile. The SNAI domain gradient also persists in the dimensions observed at municipal scale (Figure~\ref{fig:robust-radar}). These similarities indicate that the core geography of MESWI is not generated by the two supra-municipal domains.

At the same time, excluding them changes the relative position of the North and Centre. Mean MESWI becomes 101.6 in the North and 101.3 in the Centre, while the South and Islands remains markedly lower at 93.2. The inner-area mean remains below the corresponding macro-regional mean in all three areas. The revised ANOVA in Table~\ref{tab:anova-robust} attributes 29\% of variation to macro-region and 6\% to inner-area status; the interaction remains negligible. The larger inner-area effect indicates that, once the repeated provincial and regional values are removed, municipality-level remoteness becomes more salient in the remaining domains.

\begin{table}[htbp]
\centering
\caption{\textit{Robustness two-way analysis of MESWI's variance (after excluding safety and subjective well-being dimensions)}}
\label{tab:anova-robust}
\small
\resizebox{\textwidth}{!}{%
\begin{tabular}{@{}lrrrrrr@{}}
\toprule
\multicolumn{1}{c}{\textbf{Source of variation}} &
\multicolumn{1}{c}{\textbf{df}} &
\multicolumn{1}{c}{\textbf{Sum of squares}} &
\multicolumn{1}{c}{\textbf{Mean square}} &
\multicolumn{1}{c}{\textbf{$F$}} &
\multicolumn{1}{c}{\textbf{$p$ value}} &
\multicolumn{1}{c}{\textbf{Partial $\eta^2$}} \\
\midrule
Macro-region & 2 & 120,052 & 60,026 & 1,651.00 & $<0.001$ & 0.290 \\
Inner-area status & 1 & 16,846 & 16,846 & 463.34 & $<0.001$ & 0.060 \\
Macro-region $\times$ inner-area status & 2 & 285 & 143 & 3.92 & 0.020 & 0.001 \\
Residuals & 7,893 & 286,969 & 36 & -- & -- & -- \\
\bottomrule
\end{tabular}%
}
\end{table}

Most importantly, Tukey's test no longer rejects equality between North and Centre: the mean contrast is only $-0.29$ points ($p=0.376$). By contrast, the South and Islands remains 8.39 points below the North and 8.10 points below the Centre, with both contrasts significant at $p<0.001$. The two excluded domains therefore contributed disproportionately to the original North-Centre separation, but they did not create the much larger North-South divide. The robustness exercise strengthens the substantive conclusion: the dominant fault line in municipal well-being is between the Mezzogiorno and the rest of the country, while inner-area status adds a smaller but meaningful layer of disadvantage.

\subsection{Discussion}
\label{section:iv_4}

The MESWI provides a municipal-scale representation of well-being that is consistent with the conceptual breadth of BES while remaining suitable for territorial comparison. Three findings are particularly relevant to the social-indicators and place-based-policy literatures.

First, spatial scale changes what can be learned about territorial disparities. Regional averages reproduce the familiar Italian divide, but municipal maps reveal extensive heterogeneity within every region. High- and low-scoring municipalities coexist within the same administrative boundaries, cautioning against treating regional averages as sufficient evidence for the targeting of local policies. This finding is consistent with provincial and urban studies showing that finer spatial detail alters the evaluation of territorial performance \citep{Ciommi2017,Nissi2018}, and extends that insight to the full municipal universe.

The results consequently support a more differentiated approach to territorial targeting. Accessibility-based classifications remain essential for identifying municipalities facing greater costs in reaching fundamental services, but they do not fully describe the level or composition of local well-being. Municipalities within the same SNAI class may occupy very different positions in the MESWI distribution, while some municipalities outside the formal inner-area perimeter display multidimensional disadvantages comparable to those found in remote territories. The binary distinction between inner and non-inner areas should therefore be interpreted as an administrative and geographical criterion rather than as a complete measure of territorial need. Combining it with municipal well-being profiles could help identify comparable forms of disadvantage outside the SNAI perimeter and distinguish between municipalities requiring different types and intensities of intervention.

Second, the SNAI classification and the MESWI measure related but distinct constructs. SNAI is designed around the cost of reaching essential services, whereas the MESWI assesses the achieved multidimensional conditions of local communities. The pronounced service-quality gradient across the accessibility classes confirms the internal coherence of the SNAI hierarchy. At the same time, the smaller ANOVA effect associated with inner area status indicates that accessibility alone cannot account for municipal well-being. Municipalities facing similar degrees of remoteness may differ substantially in economic resources, labour-market conditions, administrative capacity, environmental quality and cultural endowments.

The macro-regional results reinforce this interpretation. Inner area status explains a non-trivial but substantially smaller share of variation than macro-region. This does not imply that remoteness is unimportant. Rather, it suggests that its effects are mediated by wider regional systems of labour demand, infrastructure, fiscal capacity, public administration, and service provision. An inner municipality located in a comparatively high-performing northern region may retain a favourable position in the national distribution, whereas a non-inner municipality in the South may still experience broad-based deprivation. Place-based programmes should therefore combine national criteria of remoteness with macro-regional and municipality-specific diagnostics.

Third, remoteness is associated with both disadvantages and amenities. Inner and peripheral municipalities perform worse in services, innovation and several socio-economic domains, but better in environmental quality. Landscape and cultural heritage display a more complex pattern, with advantages in both major hubs and some ultra-peripheral municipalities. These trade-offs matter for policy makers. Strategies focused exclusively on deficits risk overlooking assets capable of supporting local development, whereas policies centred only on amenities may underestimate the welfare costs of weak accessibility, limited employment opportunities and inadequate services.


In particular, the environmental advantage of peripheral municipalities should be regarded as a measurable territorial asset rather than as a merely symbolic compensation for economic and infrastructural disadvantages. Forests, soils, water resources, landscapes, and biodiversity provide ecosystem services whose benefits extend well beyond municipal boundaries, including carbon sequestration, hydrogeological protection, water regulation, and improvements in air and water quality \citep{Marchetti2016,MarchettiEtAl2017}. Henceforth, our results provide an empirical basis for policies that recognise the contribution of remote territories to environmental protection and to the ecological transition. Nonetheless, natural capital does not automatically generate income, employment, or demographic resilience. Depopulation and population ageing may instead weaken land management, reduce maintenance activities and increase exposure to forest fires, hydrogeological instability and environmental degradation \citep{BenassiEtAl2024}. At the same time, the productive structure of many inner areas is dominated by micro and small firms that often face limited access to finance, skills and innovation \citep{AccetturoEtAl2022,monturano2025income}. Mobilising environmental assets therefore requires institutions and productive arrangements capable of retaining locally a greater share of the value they generate. Business networks, forest agreements, green communities and renewable-energy communities may help overcome both productive and administrative fragmentation by pooling expertise, coordinating investments and connecting natural resources with local supply chains. Such forms of cooperation are especially important where individual municipalities and firms are too small to reach an efficient scale of intervention, and may generate an institutional dividend alongside their environmental and economic benefits.

Institutional capacity is therefore a central part of the policy interpretation. The weaker performance of more remote municipalities in the politics and institutions domain suggests that administrative fragility is not confined to isolated cases. This issue is particularly consequential when access to development resources depends on competitive calls, complex application procedures and the ability to design and manage projects that span over several years. Formally equal access to funding may produce unequal outcomes when the municipalities with the greatest needs possess the fewest technical and organisational resources. Place-based funding should consequently be accompanied by technical assistance, shared professional structures and support for project preparation, implementation, and monitoring. Otherwise, reliance on local planning capacity may reinforce rather than reduce existing territorial disparities.

The AMPI-based aggregation is useful in this setting because it penalises severe imbalance across domains. This is not merely a technical feature. A municipality with exceptional environmental or cultural resources but weak performance in health, education, employment, or service access should not be treated as equivalent to a municipality with the same average score and a more balanced welfare profile. The non-compensatory adjustment therefore reflects the principle that excellence in a limited number of domains cannot fully offset substantial deprivation elsewhere. At the same time, equal weighting preserves interpretability and avoids introducing preference structures that cannot be clearly justified. The robustness analysis further indicates that the principal spatial conclusions survive a consequential change in domain composition.

Several limitations qualify the analysis. First, the index is cross-sectional and combines the latest available observations from different years. Second, selected indicators require provincial or regional imputation, and the baseline results partly reflect the resulting spatial smoothing. Third, the 5\textsuperscript{th}--95\textsuperscript{th} percentile goalposts are robust for cross-sectional comparison but remain sample-dependent; fixed anchors would be necessary for rigorous inter-temporal monitoring. Fourth, equal municipal weighting gives small and large municipalities the same influence in group means. This is appropriate when municipalities are the units of interest, but it does not reproduce the distribution of well-being among residents. Fifth, the analysis does not explicitly model spatial autocorrelation, measurement uncertainty, or the mechanisms through which remoteness affects different well-being domains.

Finally, ANOVA $p$ values should not be over-interpreted because the study covers nearly the entire municipal population and group variances are unequal. Mean contrasts, effect sizes and mapped distributions provide more informative evidence of substantive territorial differences. The factorial model should therefore be read as a descriptive, model-based decomposition of observed variability rather than as a causal design.

These limitations point to several directions for further research. A panel version of the MESWI could use fixed goalposts to evaluate territorial convergence, decline and exposure to place-based interventions. Population-weighted and resident-level measures could complement the municipal-unit perspective. Spatial econometric models could identify clusters and spillovers, while uncertainty analysis could assess sensitivity to imputation, domain composition and alternative weighting schemes. Participatory weighting would also allow locally expressed priorities to be compared with the equal-weight benchmark \citep{ResceMaynard2018,Greco2020,Liberati2022}. Further improvements in the availability of municipal data on administrative capacity and social inclusion would also broaden the range of territorial inequalities captured by the index.

\section{Conclusions}
\label{section:v} 

This paper develops the Municipal Equitable and Sustainable Well-being Index (MESWI), a nationwide measure of multidimensional well-being covering all 7,896 Italian municipalities. By translating the 12-domain BES (\textit{Benessere Equo e Sostenibile}) framework to the municipal scale through 49 statistical and administrative indicators, the MESWI brings together two aspects of well-being measurement that are often treated separately: multidimensionality and territorial granularity. The hierarchical aggregation based on the Adjusted Mazziotta--Pareto Index preserves the domain structure of the BES while penalising markedly unbalanced well-being profiles.

Results show that the familiar North-South divide remains clearly visible and represents the dominant geographical pattern in Italian well-being, outweighing the distinction between Inner and non-Inner Areas. Indeed, an Inner-Area municipality in the North often records higher levels of multidimensional well-being than a more accessible municipality in the South. At the same time, municipal data reveal substantial variation within regions and broader territorial categories, with municipalities located in the same regional context displaying markedly different levels and compositions of well-being. Aggregate territorial statistics therefore provide an important but incomplete picture of the geography of well-being in Italy.

The analysis of inner areas provides a particularly informative illustration of this heterogeneity. Remoteness is associated on average with weaker performance in services, innovation and several socio-economic dimensions, but the relationship is neither uniform nor exclusively negative. Peripheral municipalities also display comparatively favourable environmental conditions and, in some cases, important cultural, and landscape endowments. More generally, municipalities facing similar geographical and accessibility conditions can differ considerably in their overall well-being and in the combination of strengths and weaknesses underlying it. The evidence therefore reinforces the need to observe territorial conditions at the scale at which many of these differences actually emerge.

This has implications for public policy that extend beyond any individual territorial programme. Policies are increasingly designed and implemented at fine geographical scales, while the information available for their programming and assessment often remains more aggregated. Municipal multidimensional indicators can help narrow this informational gap. The MESWI can provide a common framework for describing local conditions, identifying concentrations of disadvantage, recognising territorial assets and comparing municipalities across a consistent set of well-being dimensions. Its domain structure is particularly useful in this respect, since similar aggregate scores may conceal very different local profiles.

The index provides a detailed cross-sectional diagnosis of municipal well-being. With regular updating and a consistent inter-temporal structure, however, the same framework could support different stages of the public-policy cycle. Before an intervention, it could contribute to the identification of territorial needs and baseline conditions; during implementation, it could provide a common set of indicators for monitoring local trajectories; and over a longer horizon, it could contribute to the assessment of whether gaps in specific dimensions of well-being are narrowing or persisting. Establishing causal policy effects would still require appropriate evaluation designs, but a stable municipal information system would provide an important empirical basis for such analyses.

The policy relevance of municipal measurement also lies in what it reveals about the composition of territorial development. Local disadvantage rarely consists of a single deficit. Weak employment opportunities may coexist with good environmental conditions; limited accessibility may coexist with significant cultural resources; and similar levels of overall well-being may result from very different combinations of domain-specific outcomes. Looking below the aggregate score is therefore essential. 

The broader contribution of the paper is the provision of a finer informational scale for the study of well-being. Moving beyond GDP broadens what is measured; moving to the municipal level changes where those differences can be observed. Taken together, these two shifts reveal a geography of Italian well-being that is more heterogeneous than regional averages suggest. As municipal data improve and longer time series become available, the MESWI can develop from a cross-sectional measurement framework into a tool for the programming, monitoring and evaluation of public policies at the territorial scale at which their effects are ultimately experienced.

\pagebreak

\bibliographystyle{apalike-etal-in-italics} 
\bibliography{bib}



\pagebreak

\section*{Appendix}
\label{section:appendix} 

\setcounter{section}{0}
\setcounter{subsection}{0}
\setcounter{subsubsection}{0}
\renewcommand{\thesubsection}{A.\arabic{subsection}}
\renewcommand{\thesubsubsection}{A.\arabic{subsection}.\arabic{subsubsection}}
\setcounter{table}{0}
\renewcommand{\thetable}{A\arabic{table}}
\setcounter{figure}{0}
\renewcommand{\thefigure}{A\arabic{figure}}

\subsection{Distance from essential services}
\label{subsection:appendix_1}

This Appendix describes the construction of the three travel-time indicators included in the \textit{Quality of services} domain. Each indicator measures the average driving time from a municipality to the nearest relevant essential service, as defined within the Italian National Strategy for Inner Areas (\textit{Strategia Nazionale per le Aree Interne} -- SNAI).

The three essential services are defined as follows:
\begin{enumerate}[label=\roman*)]
\item \textit{education}: the availability of a diversified upper-secondary education supply, comprising at least one general high school and one technical or vocational institute;
\item \textit{healthcare}: the presence of a hospital equipped with an emergency department and capable of managing complex emergencies, including resuscitation services, corresponding to an emergency and admission department of at least first level;\footnote{\textit{Dipartimento d'Emergenza e Accettazione} (DEA).}
\item \textit{mobility}: the presence of a railway station classified at least as silver category.\footnote{This category includes two types of facilities: i) medium-sized or small stations and stops with substantial passenger volumes---approximately more than 2,500 passengers per day---and services supporting both long- and short-distance travel; and ii) medium-sized or small stations and stops with substantial or high passenger volumes within urban and metropolitan railway systems---in some cases exceeding 4,000 passengers per day---which are often unstaffed, lack a passenger building (\textit{fabbricato viaggiatori}), and are served exclusively by regional or metropolitan trains.}
\end{enumerate}

In the original SNAI methodology, municipalities were classified according to the driving time to the nearest service hub in which all three essential services were jointly available \citep{nuvap2020strategia}. Hubs generally corresponded to provincial capitals or to groups of neighbouring municipalities that collectively provided the required education, healthcare, and mobility services. Travel times were calculated between municipal centroids using \href{https://docs.tomtom.com/}{TomTom} data collected daily between October 14 and October 20, 2019. The work of \cite{bergantino2026far} proposed an updated classification of inner areas trying to replicate the \citeauthor{nuvap2020strategia}'s approach. 

The indicators proposed in this study build on this approach but extend it in two respects. First, travel times are calculated for all Italian municipalities rather than only for those located outside service hubs. Second, accessibility is assessed separately for each essential service. We therefore geocoded the locations of general high schools and technical or vocational institutes \citep{mim_unica}, railway stations classified at least as silver category \citep{rfi_stazioni}, and hospitals with a DEA of at least first level \citep{msal_asl}. For each municipality, the geographical coordinates of the town hall were used as the point of origin \citep{garda_info}. Town halls were preferred to geometric centroids because they generally provide a more representative approximation of the historical and administrative centre around which municipal settlements developed.

Using the \href{https://www.openstreetmap.org/#map=6/42.09/12.56}{OpenStreetMap} (OSM) road network, we first constructed origin--destination driving-time matrices between each town hall and the relevant service facilities. For each municipality and service type, the set of potential destinations included all facilities located within a radius of 5 kilometres. Where no facility was located within this radius, the facility with the shortest estimated driving time in the OSM matrix was selected.\footnote{For the Metropolitan City of Rome, travel times were calculated separately for each of its fifteen municipal districts (\textit{municipi}) and subsequently averaged.}

The resulting origin--destination subsets were then submitted to the TomTom routing service to obtain traffic-sensitive driving times. For education, driving times were calculated separately for general high schools and technical or vocational institutes; the education indicator was subsequently obtained by averaging the results for the two types of institution. For each municipality--service combination, the final value corresponds to the average of three daily routing queries, initiated at 7:00, 8:00, and 9:00 a.m., on each working day between October 20 and October 24, 2025.\footnote{The developer API does not support bulk downloads. Route queries---each corresponding to a request between one origin and one destination---were therefore submitted sequentially, with short delays introduced to reduce the likelihood of response errors. Each complete set of queries, covering all municipalities rather than only SNAI hubs, required approximately 8-9 hours. To prevent municipalities from being systematically queried at the same time of day, their order was randomly reshuffled for each set of requests. The analysis was conducted entirely in \href{https://www.r-project.org/}{R}, using the packages \texttt{osrm} and \texttt{tomtom} to access OSM and TomTom data, respectively.}

\begin{figure}[H]
\centering
\caption{\textit{Workflow for calculating driving times to essential services and assigning Inner Areas categories}}
\includegraphics[width=0.94\textwidth]{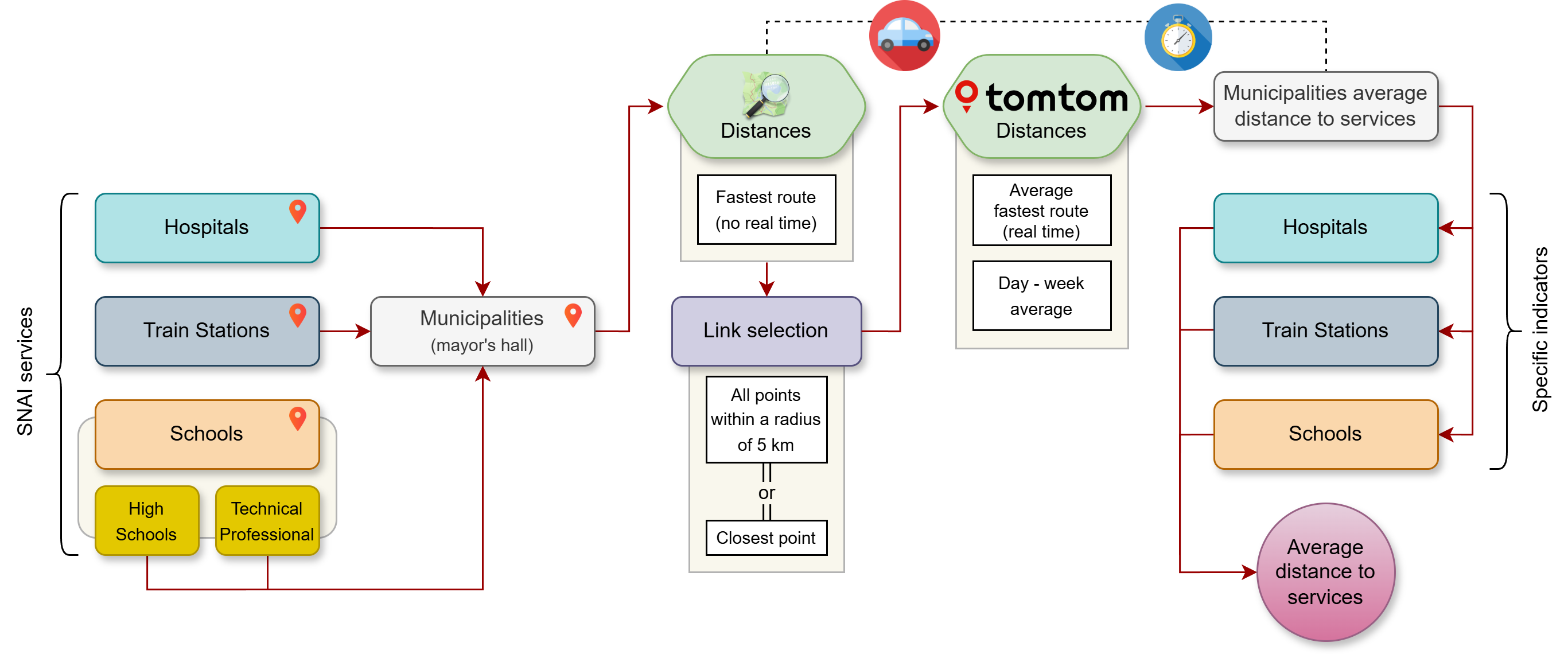}
\label{fig:workflow}
\end{figure}

\begin{figure}[H]
    \centering
    \caption{\textit{Service-specific inner areas classification (2025)}}
    \vspace{-.2cm}
    \begin{minipage}[b]{0.32\linewidth}
        \centering
        \footnotesize \textbf{Schools} \\
        \tiny Inner Areas: 992 (12.5\%) \\
        \tiny Average time: 18.1 minutes
        \vspace{-.5cm}
    \end{minipage}
    \begin{minipage}[b]{0.32\linewidth}
        \centering
        \footnotesize \textbf{Railway stations} \\
        \tiny Inner Areas: 2,917 (36.9\%) \\
        \tiny Average time: 26.6 minutes
        \vspace{-.5cm}
    \end{minipage}
    \hfill
    \begin{minipage}[b]{0.32\linewidth}
        \centering
        \footnotesize \textbf{Hospitals} \\
        \tiny Inner Areas: 3,678 (46.5\%) \\
        \tiny Average time: 30.1 minutes
        \vspace{-.5cm}
    \end{minipage}
    \includegraphics[height=0.6\linewidth]{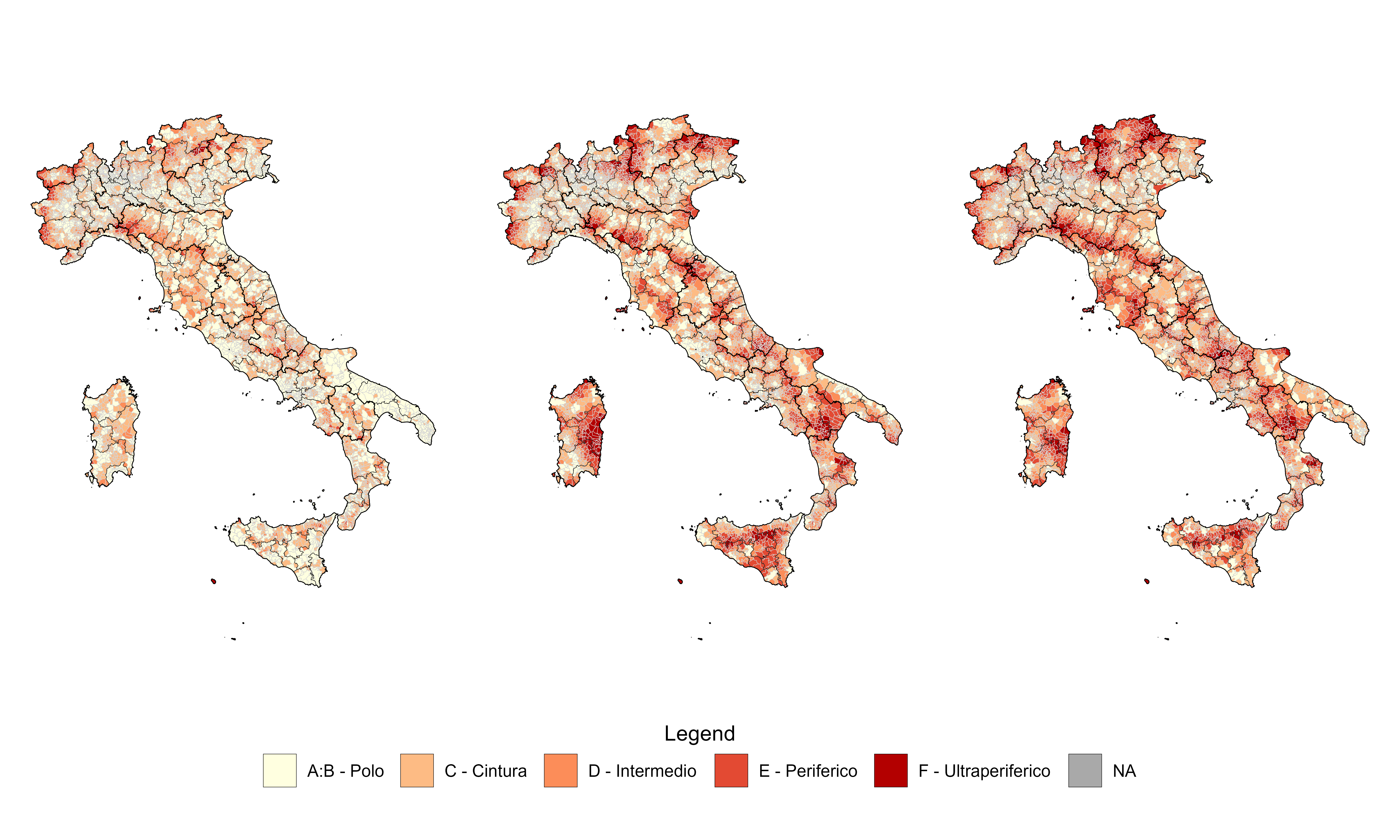}
    \label{fig:snai_dis_ser}
    \begin{minipage}{0.98\linewidth}
    \vspace{-.5cm}
    \justifying
    \tiny
    \noindent
    Notes: TomTom data were collected daily between October 20 and October 24, 2025. Municipalities with a driving time below 15 minutes were classified as \textit{A--B: hubs} \citep{moreno2021introducing}.\\
    Sources: authors' elaboration based on \cite{msal_asl,garda_info,istat_confini,mim_unica,rfi_stazioni}.
    \end{minipage}
\end{figure}

The methodological workflow is summarised in Figure~\ref{fig:workflow}. Figure~\ref{fig:snai_dis_ser} presents a service-specific classification of Italian municipalities based on the resulting driving times. To facilitate comparison with the official SNAI classification for 2021--2027, we applied the same thresholds \citep{nuvap2020strategia}: municipalities with driving times below 27.7 minutes were classified as belt areas; those with driving times from 27.7 to less than 40.9 minutes as intermediate areas; those with driving times from 40.9 to less than 66.9 minutes as peripheral areas; and those with driving times of at least 66.9 minutes as ultra-peripheral areas.

The data were calculated by \cite{InstantAnalytics2025}.

\subsection{Figures and Tables}
\label{subsection:appendix_2}




\begin{figure}[H]
\centering
\caption{\textit{MESWI deciles and SNAI project areas for each region}}
\label{fig:regions-map}

\begin{tabular}{
    >{\centering\arraybackslash}p{0.32\textwidth}
    >{\centering\arraybackslash}p{0.32\textwidth}
    >{\centering\arraybackslash}p{0.32\textwidth}
}

{\small\textbf{Piedmont}} &
{\small\textbf{Aosta Valley}} &
{\small\textbf{Lombardy}}
\\[-1mm]

\includegraphics[width=\linewidth]{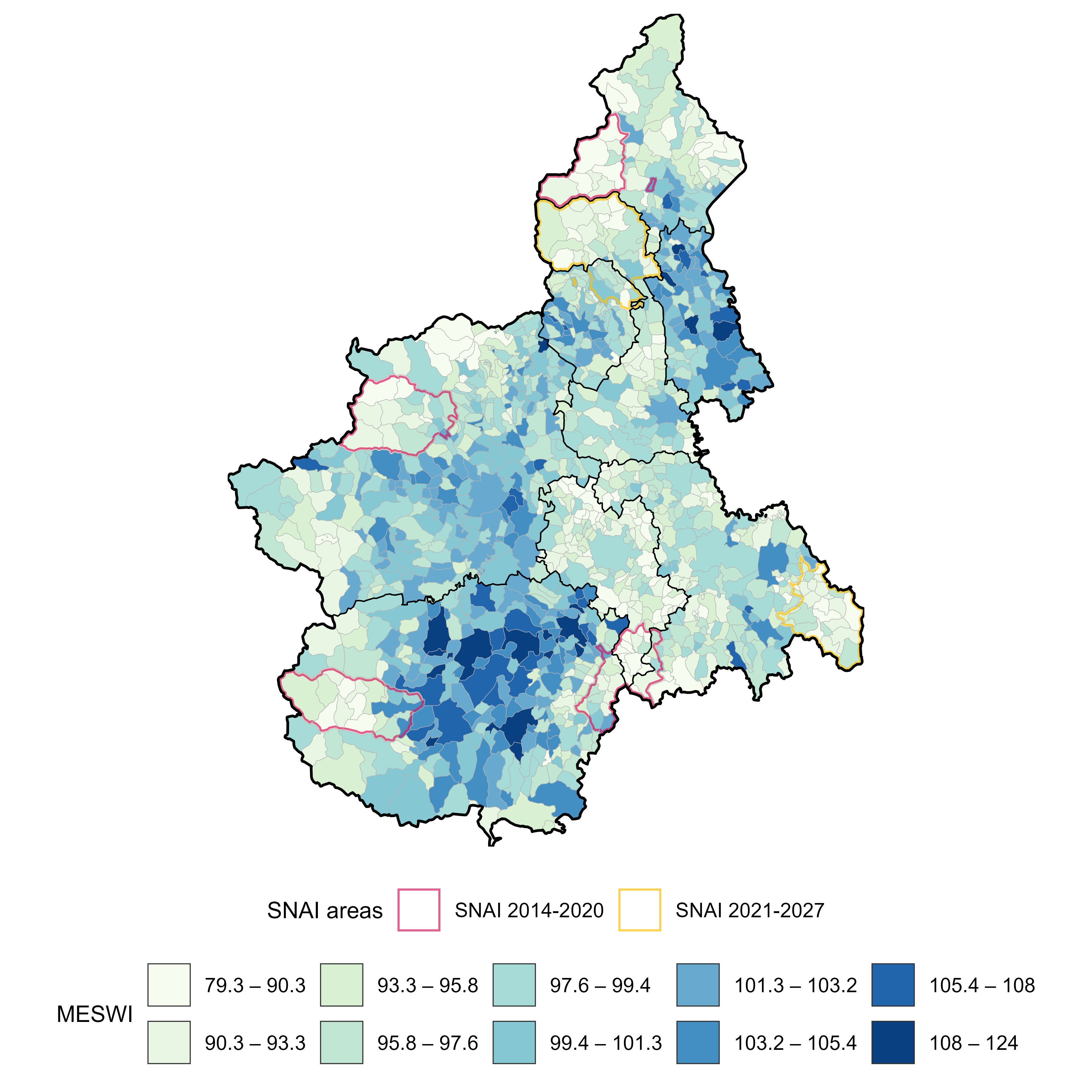} &
\includegraphics[width=\linewidth]{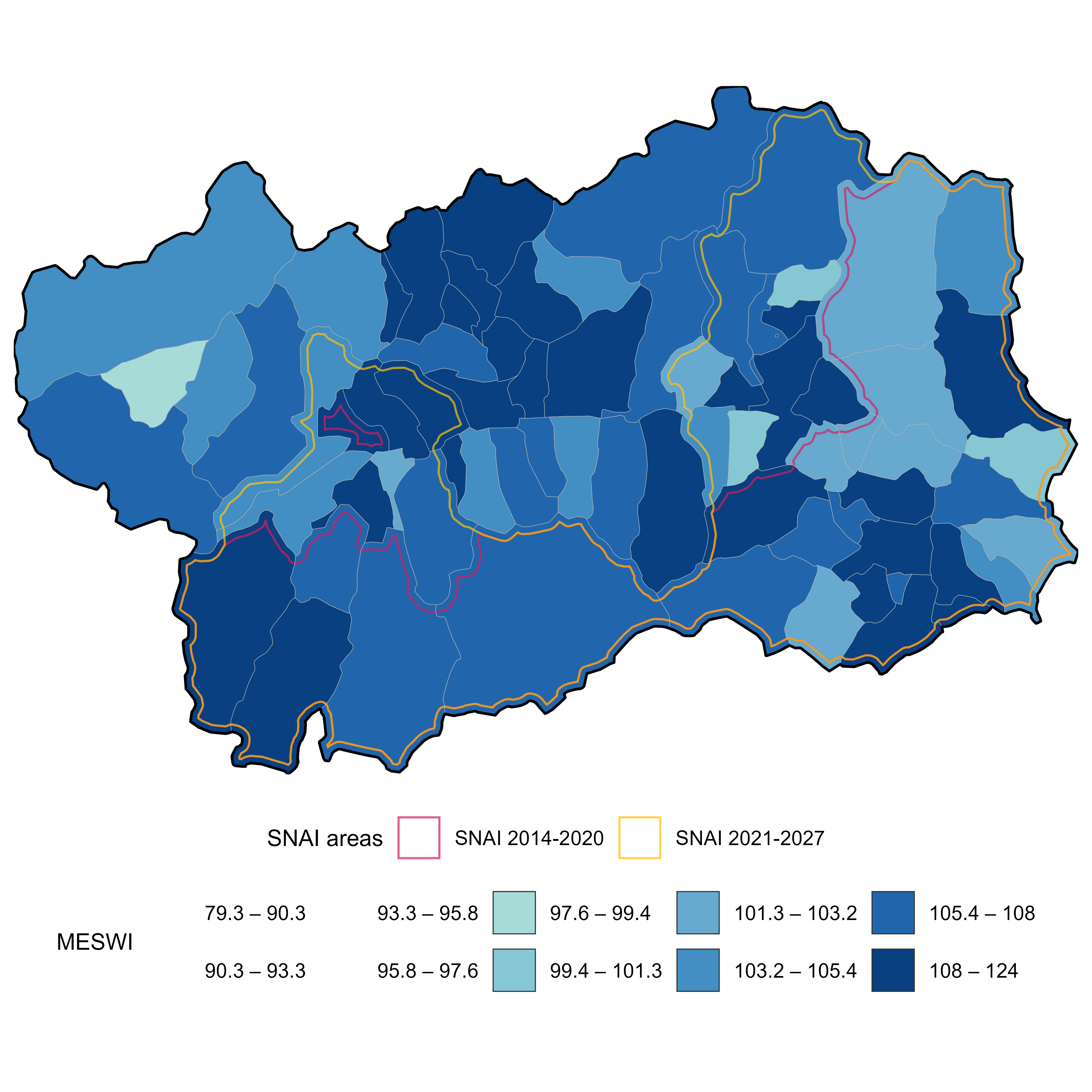} &
\includegraphics[width=\linewidth]{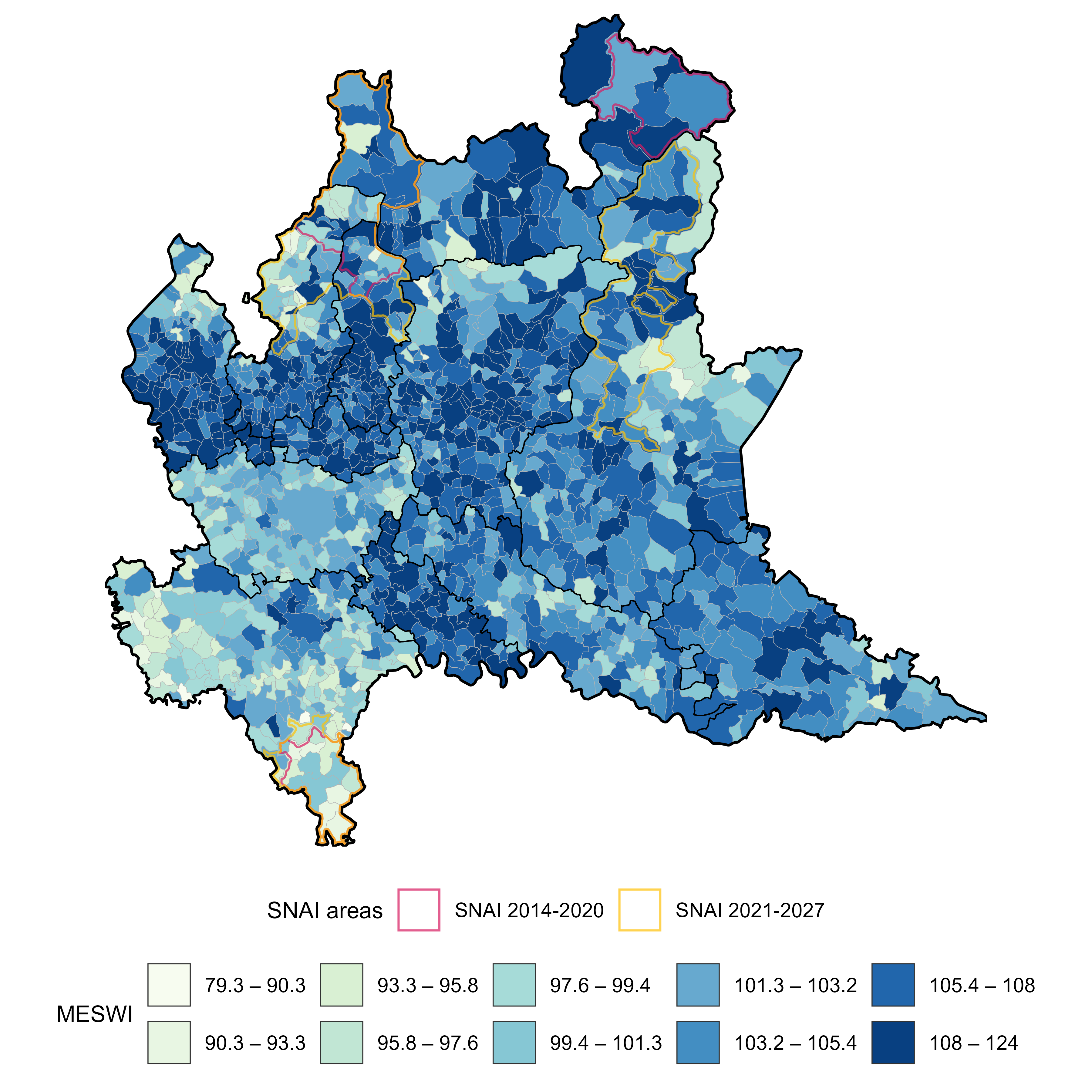}
\\[2mm]

{\small\textbf{Trentino-South Tyrol}} &
{\small\textbf{Veneto}} &
{\small\textbf{Friuli-Venezia Giulia}}
\\[-1mm]

\includegraphics[width=\linewidth]{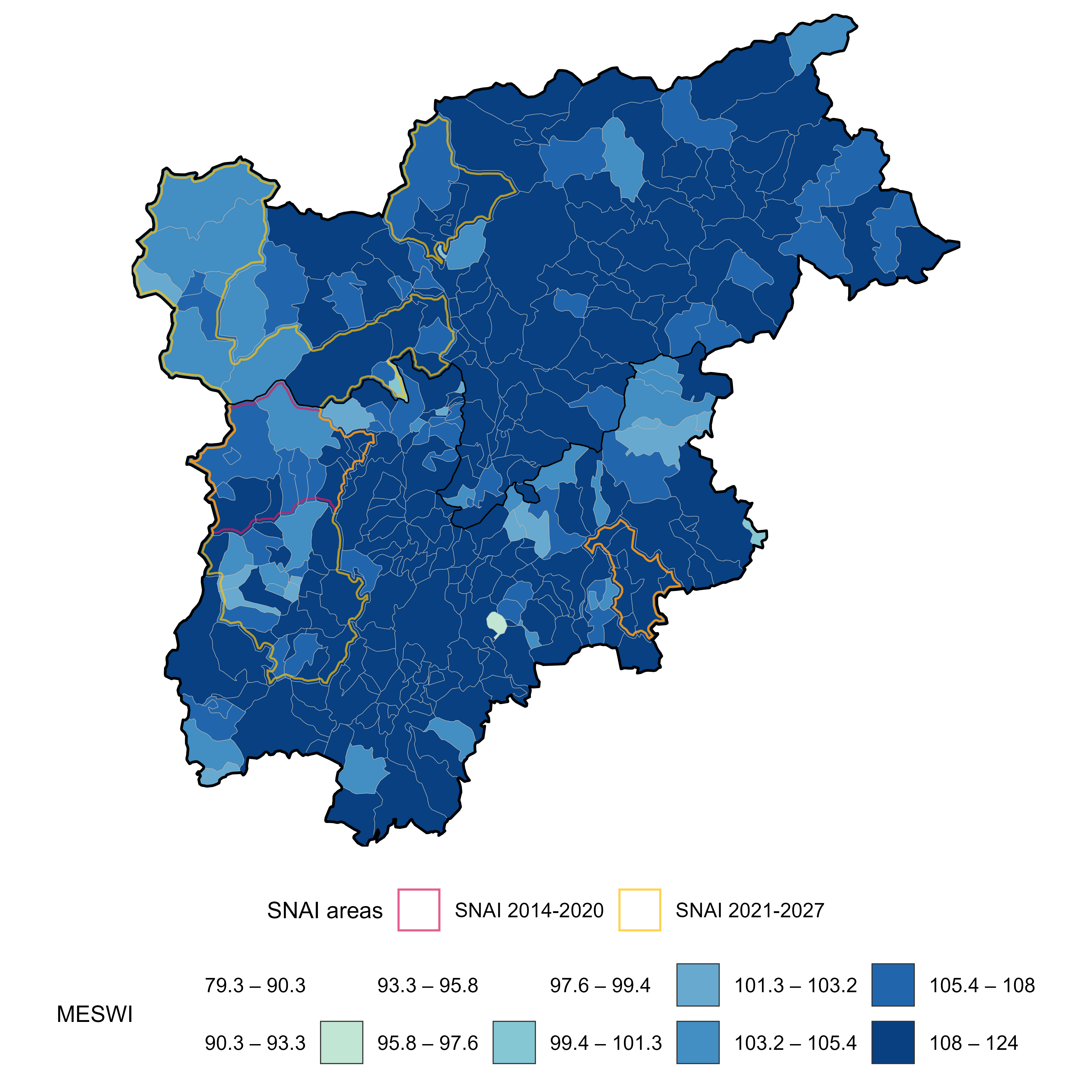} &
\includegraphics[width=\linewidth]{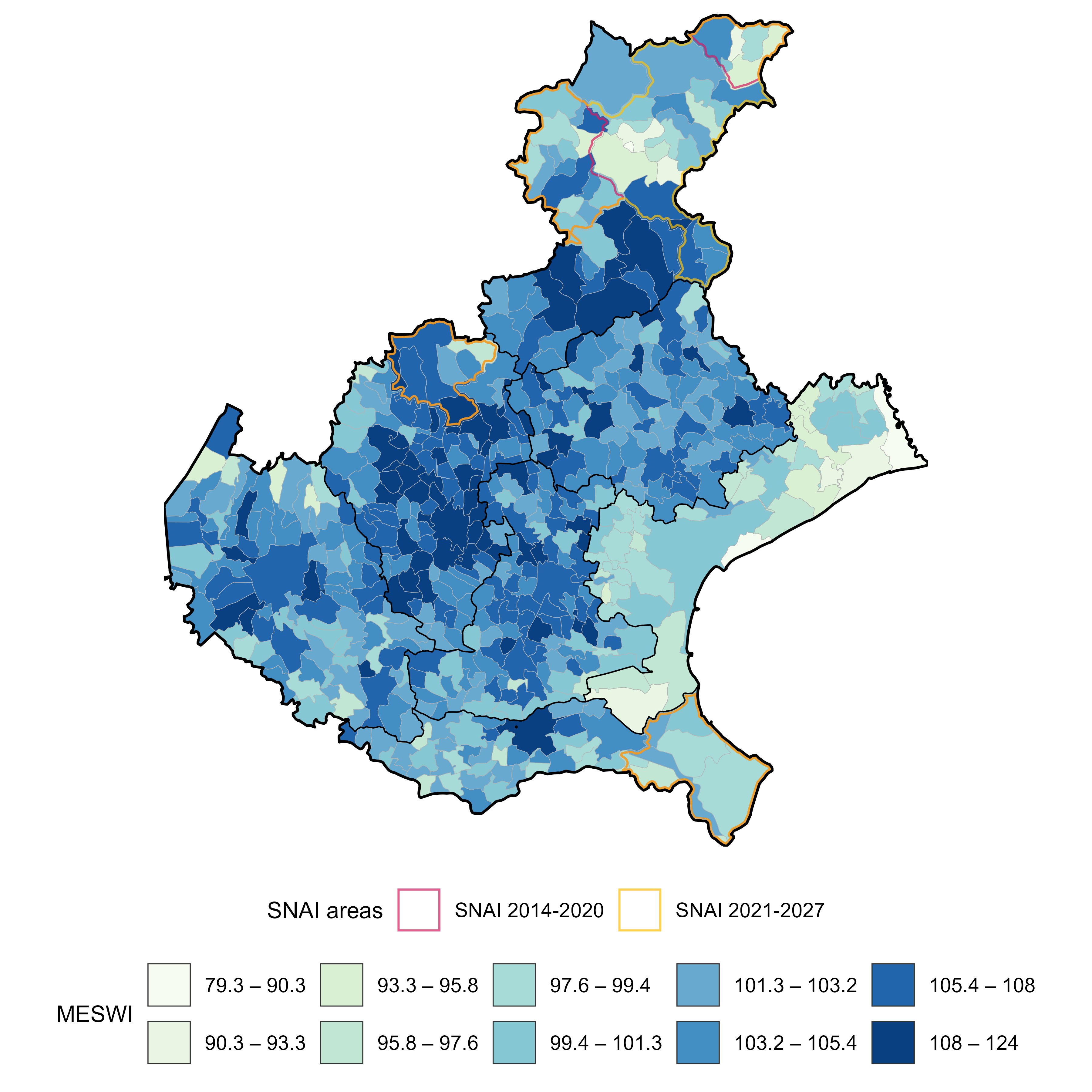} &
\includegraphics[width=\linewidth]{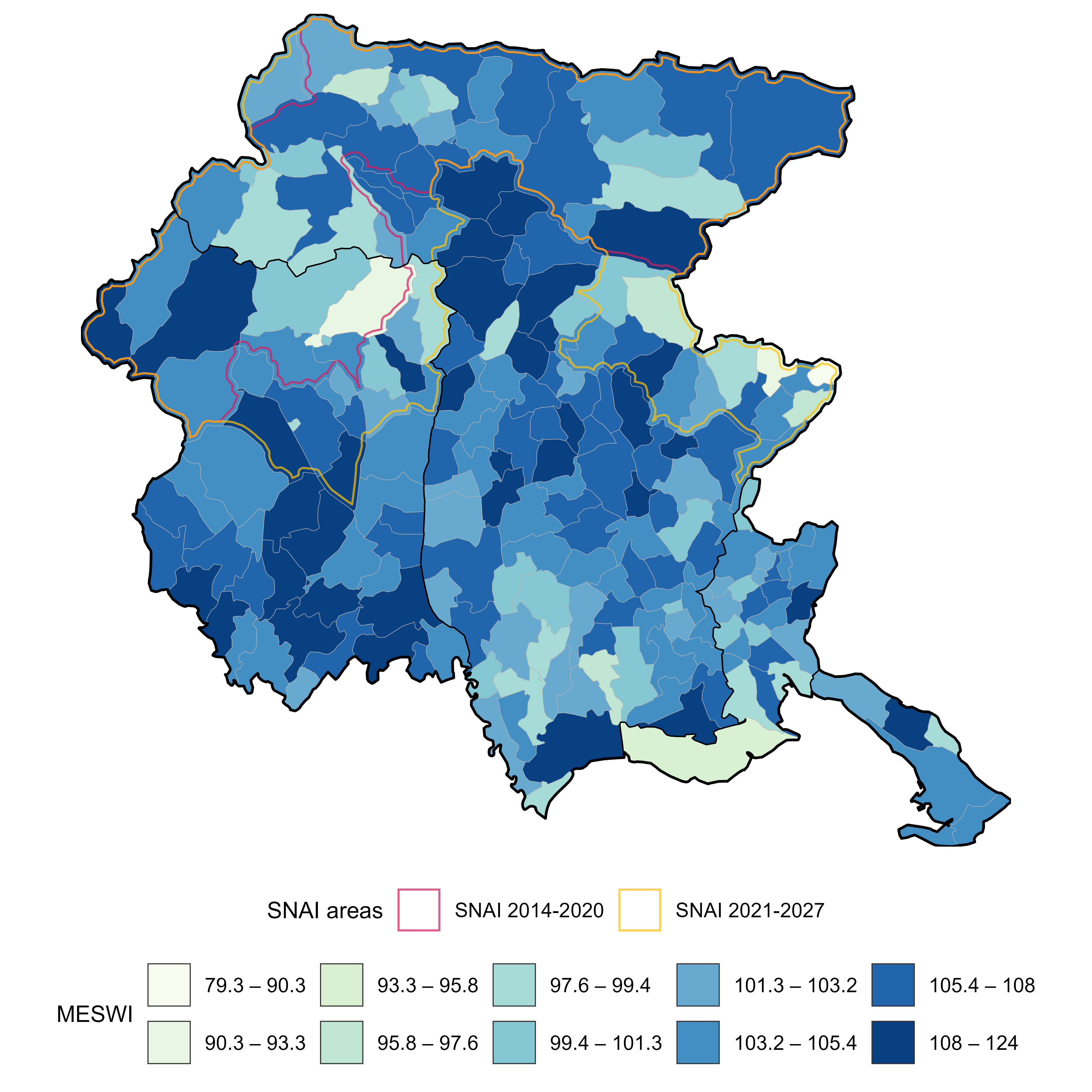}
\\[2mm]

{\small\textbf{Liguria}} &
{\small\textbf{Emilia-Romagna}} &
{\small\textbf{Tuscany}}
\\[-1mm]

\includegraphics[width=\linewidth]{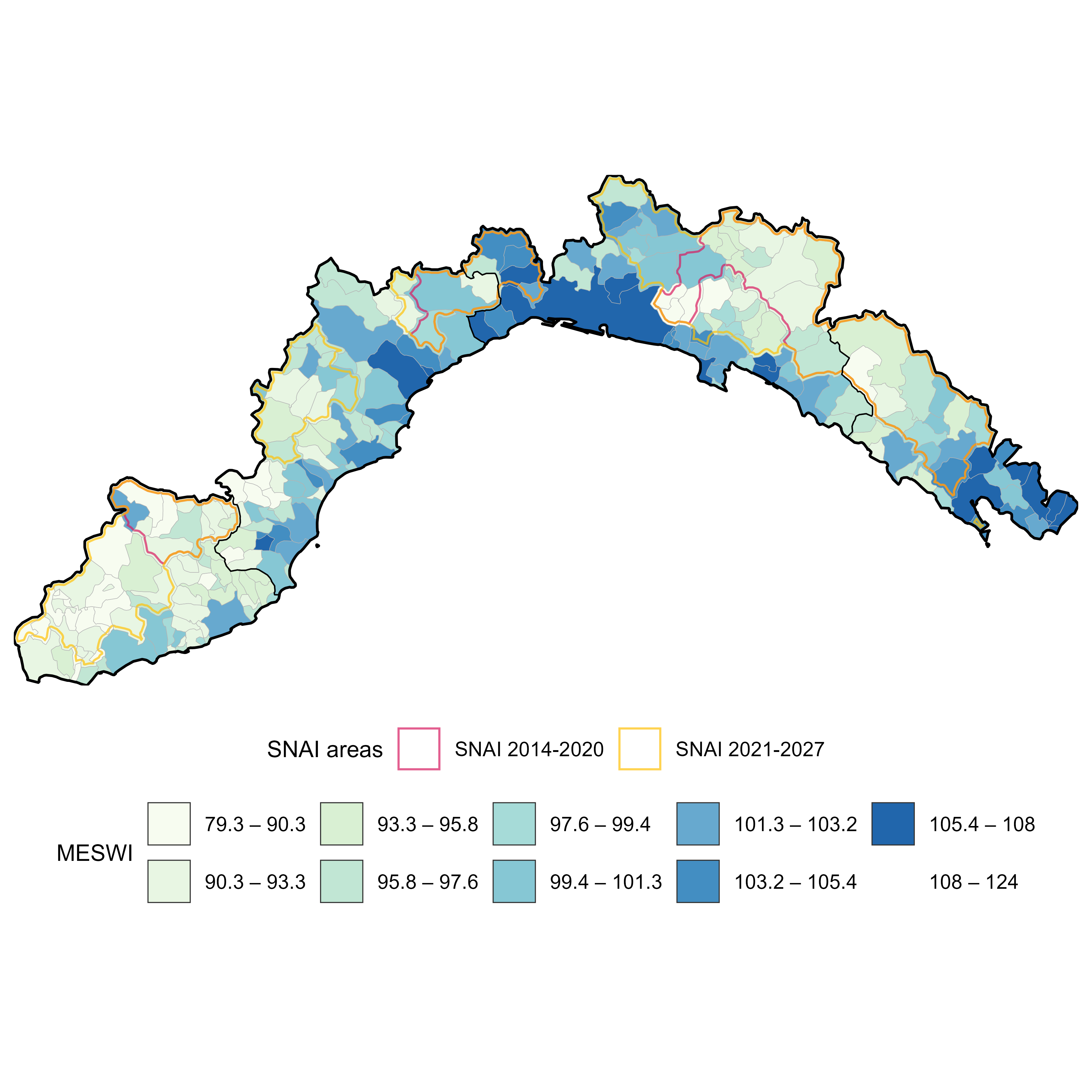} &
\includegraphics[width=\linewidth]{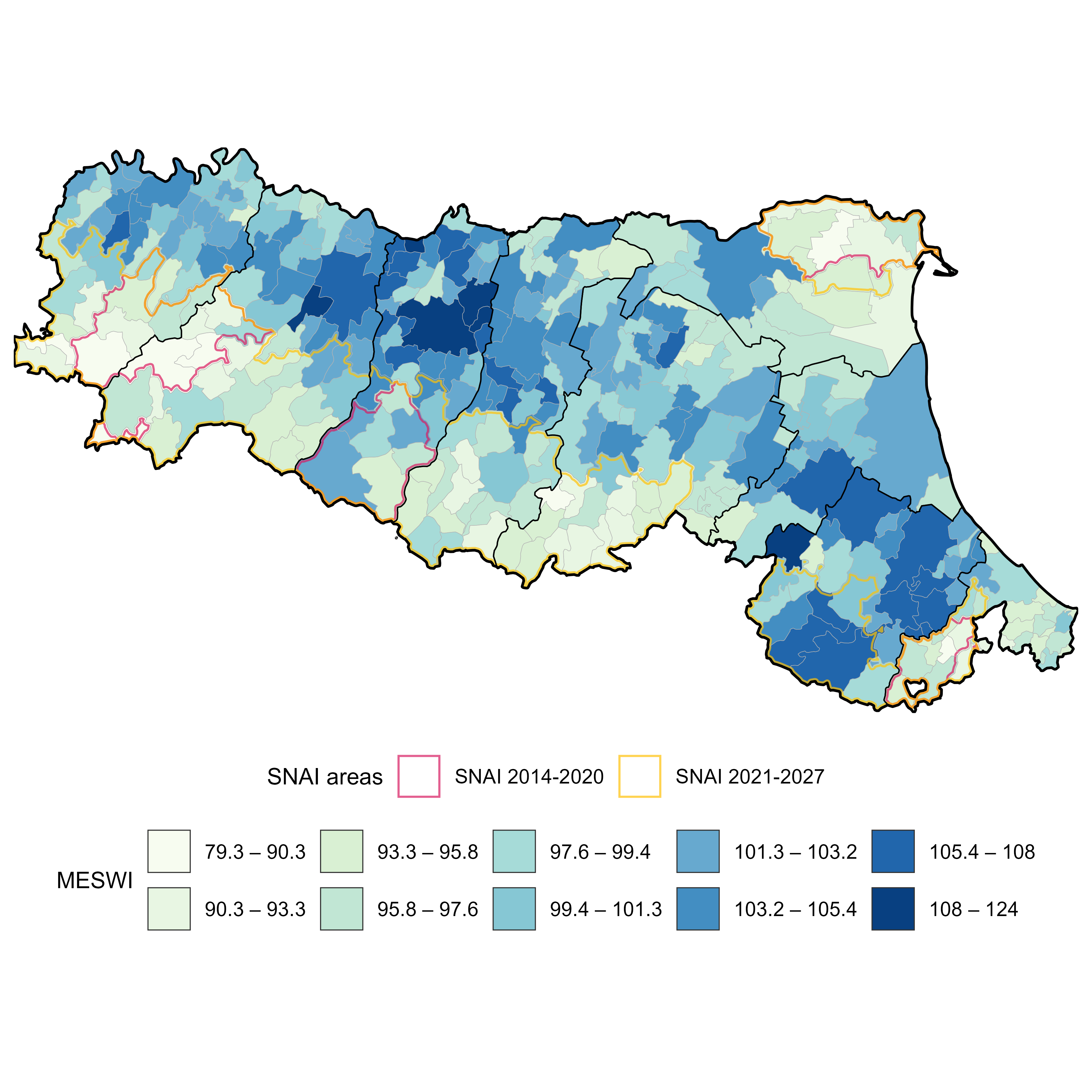} &
\includegraphics[width=\linewidth]{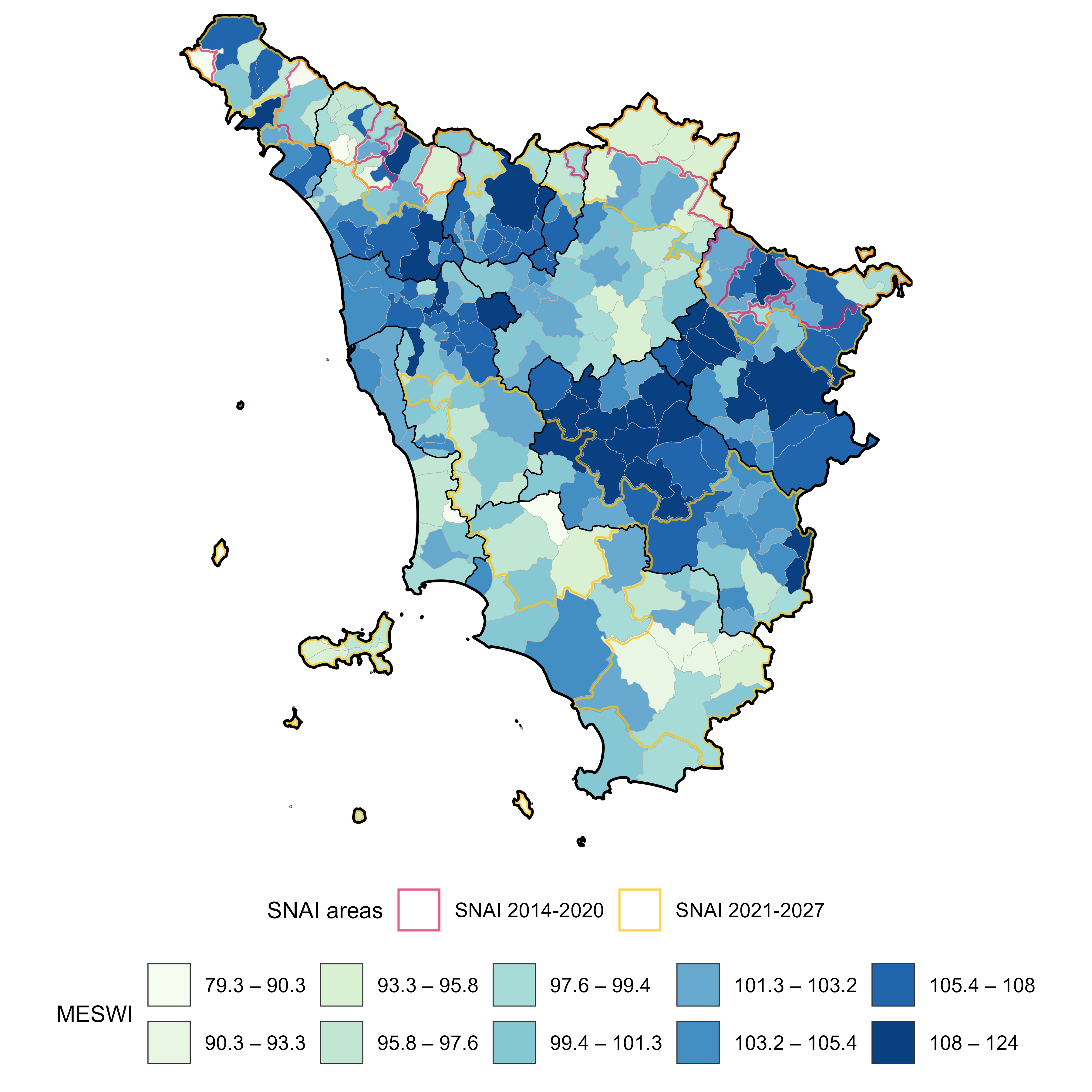}

\end{tabular}
\end{figure}


\begin{figure}[H]
\centering
\ContinuedFloat
{\small\textit{(continued)}}

\begin{tabular}{
    >{\centering\arraybackslash}p{0.32\textwidth}
    >{\centering\arraybackslash}p{0.32\textwidth}
    >{\centering\arraybackslash}p{0.32\textwidth}
}

{\small\textbf{Umbria}} &
{\small\textbf{Marche}} &
{\small\textbf{Lazio}}
\\[-1mm]

\includegraphics[width=\linewidth]{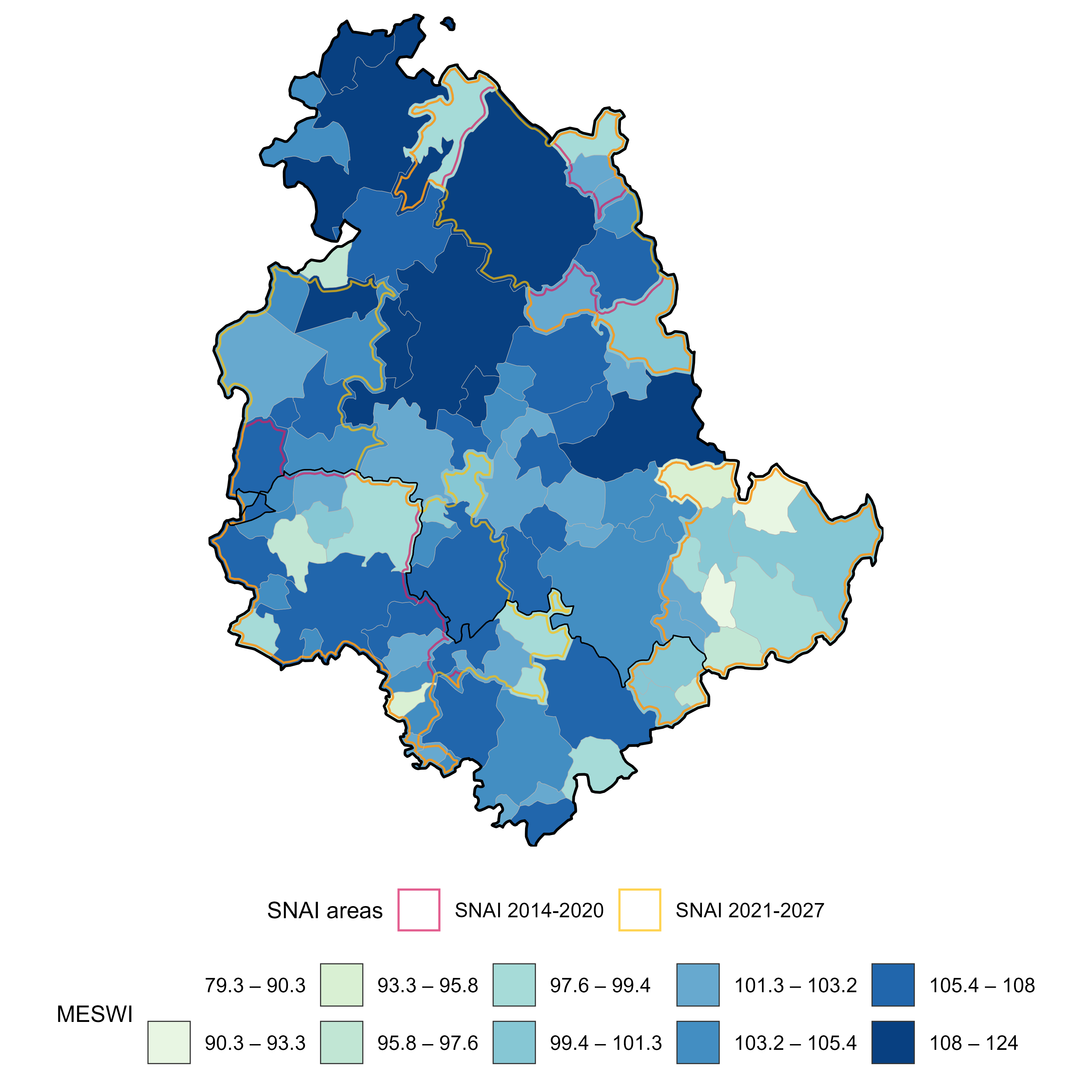} &
\includegraphics[width=\linewidth]{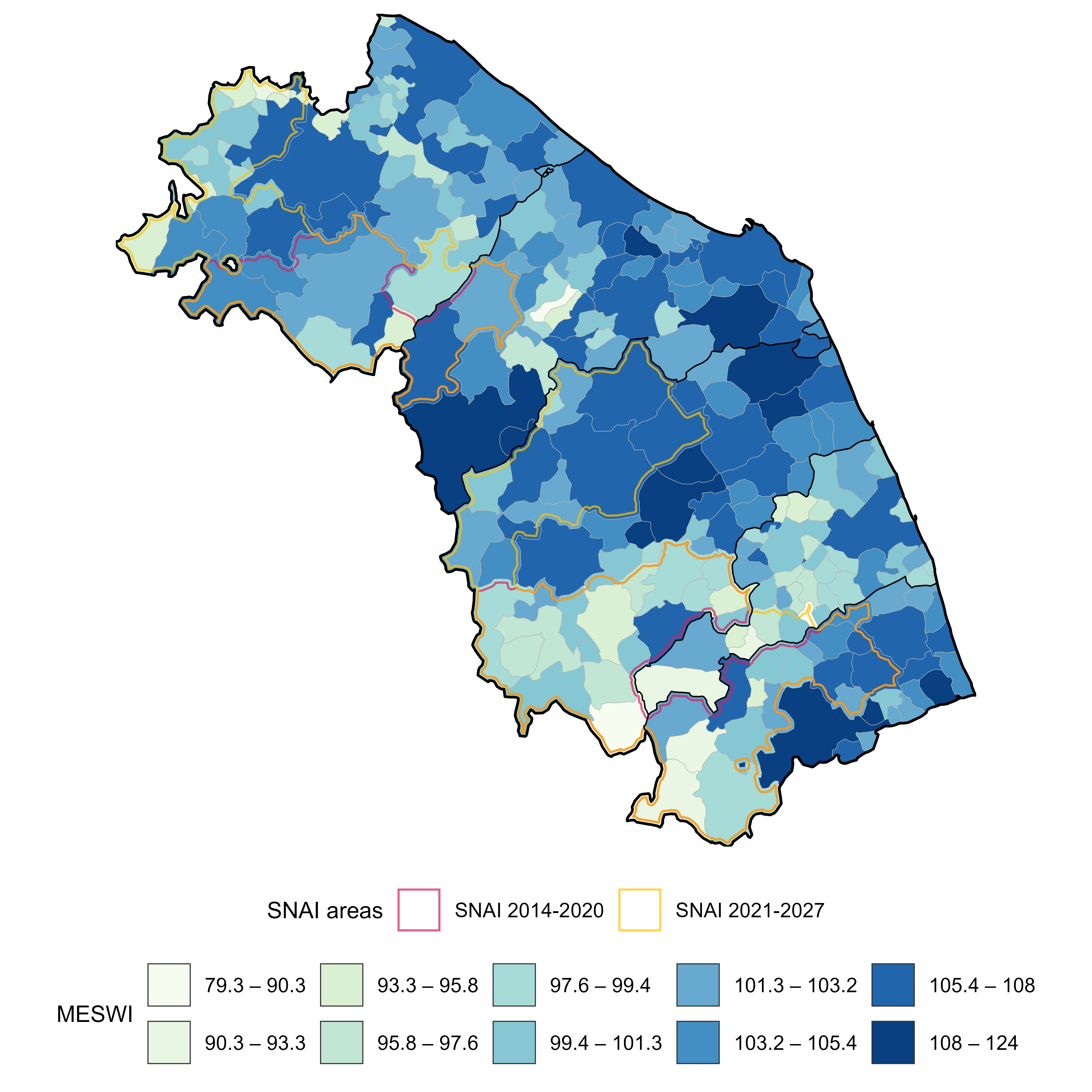} &
\includegraphics[width=\linewidth]{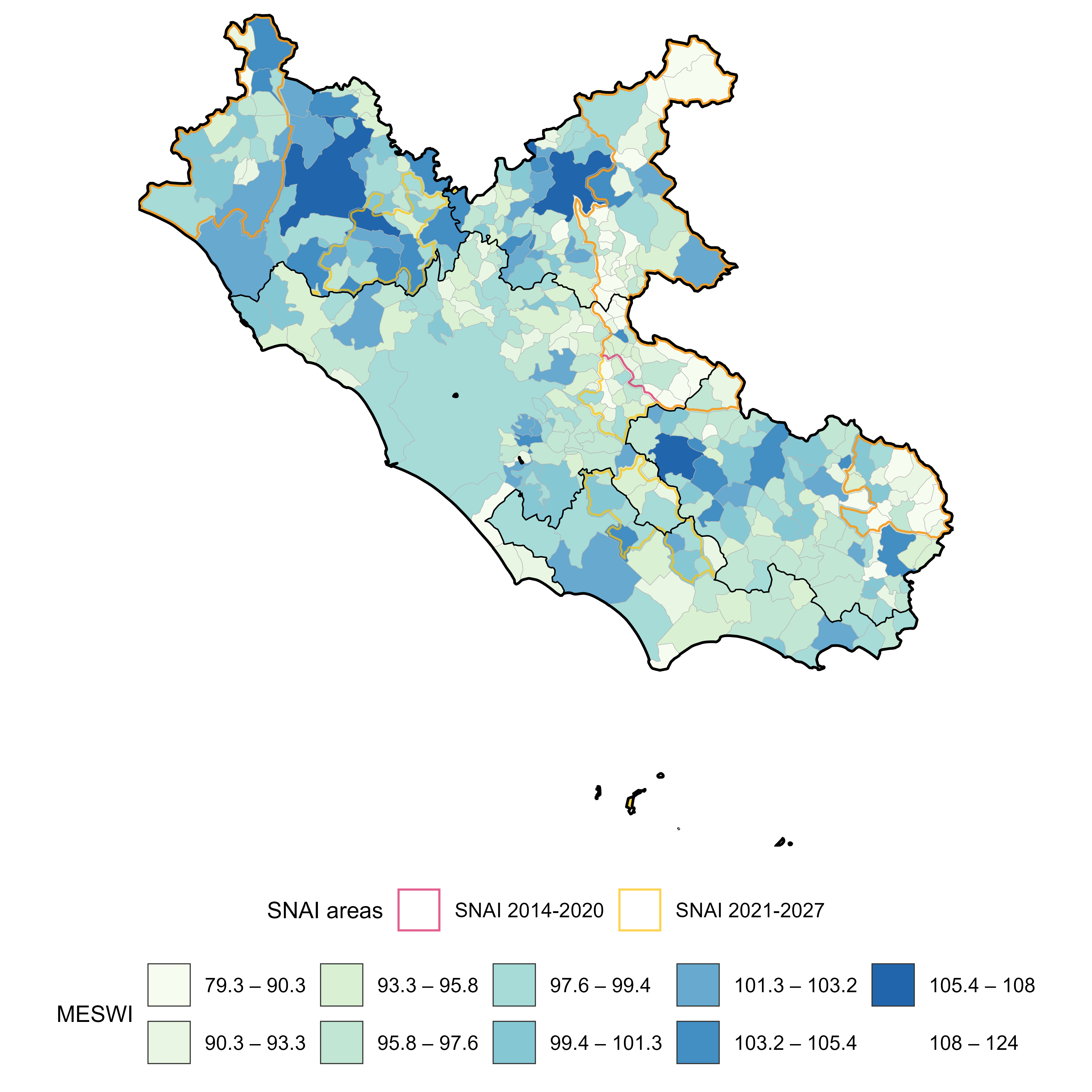}
\\[2mm]

{\small\textbf{Abruzzo}} &
{\small\textbf{Molise}} &
{\small\textbf{Campania}}
\\[-1mm]

\includegraphics[width=\linewidth]{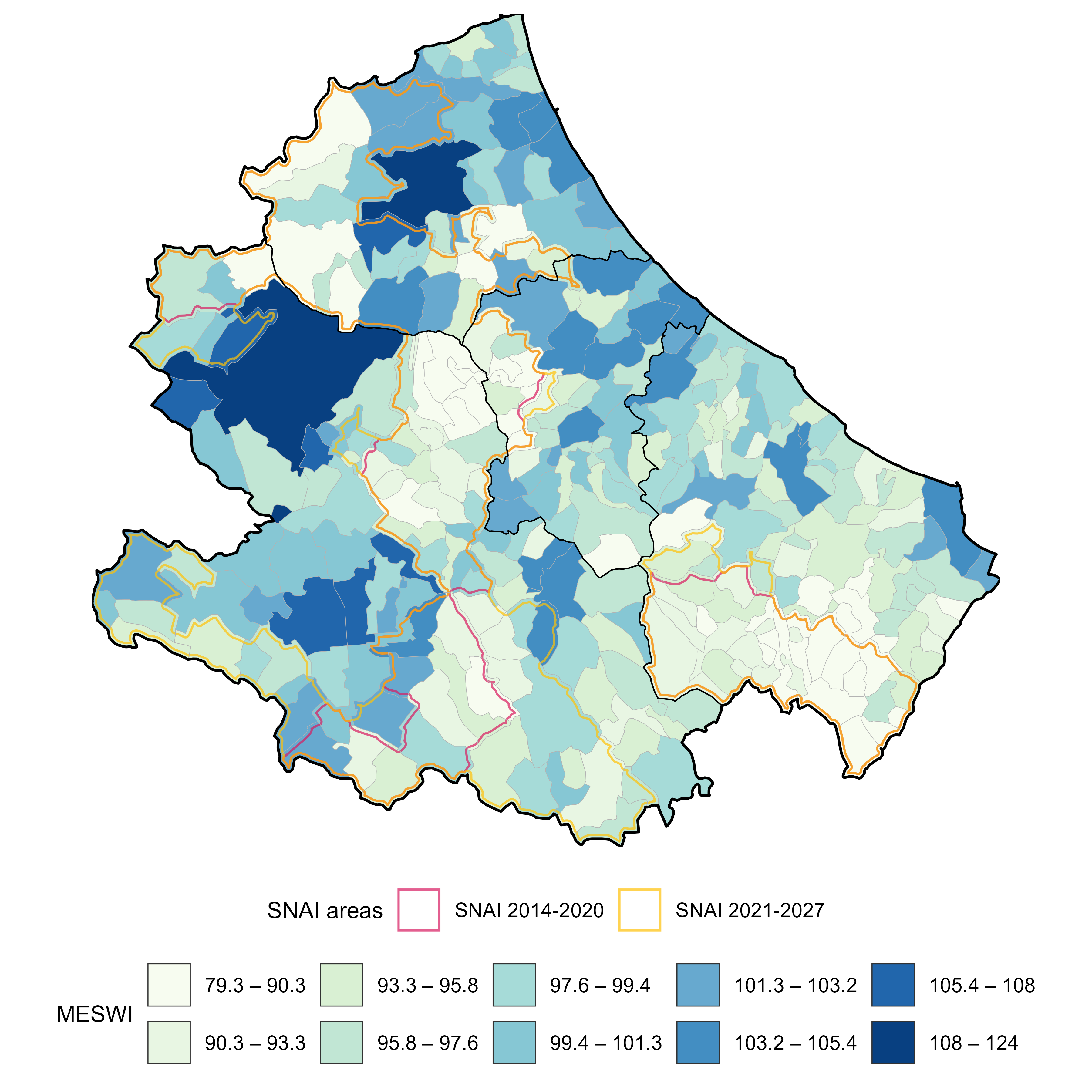} &
\includegraphics[width=\linewidth]{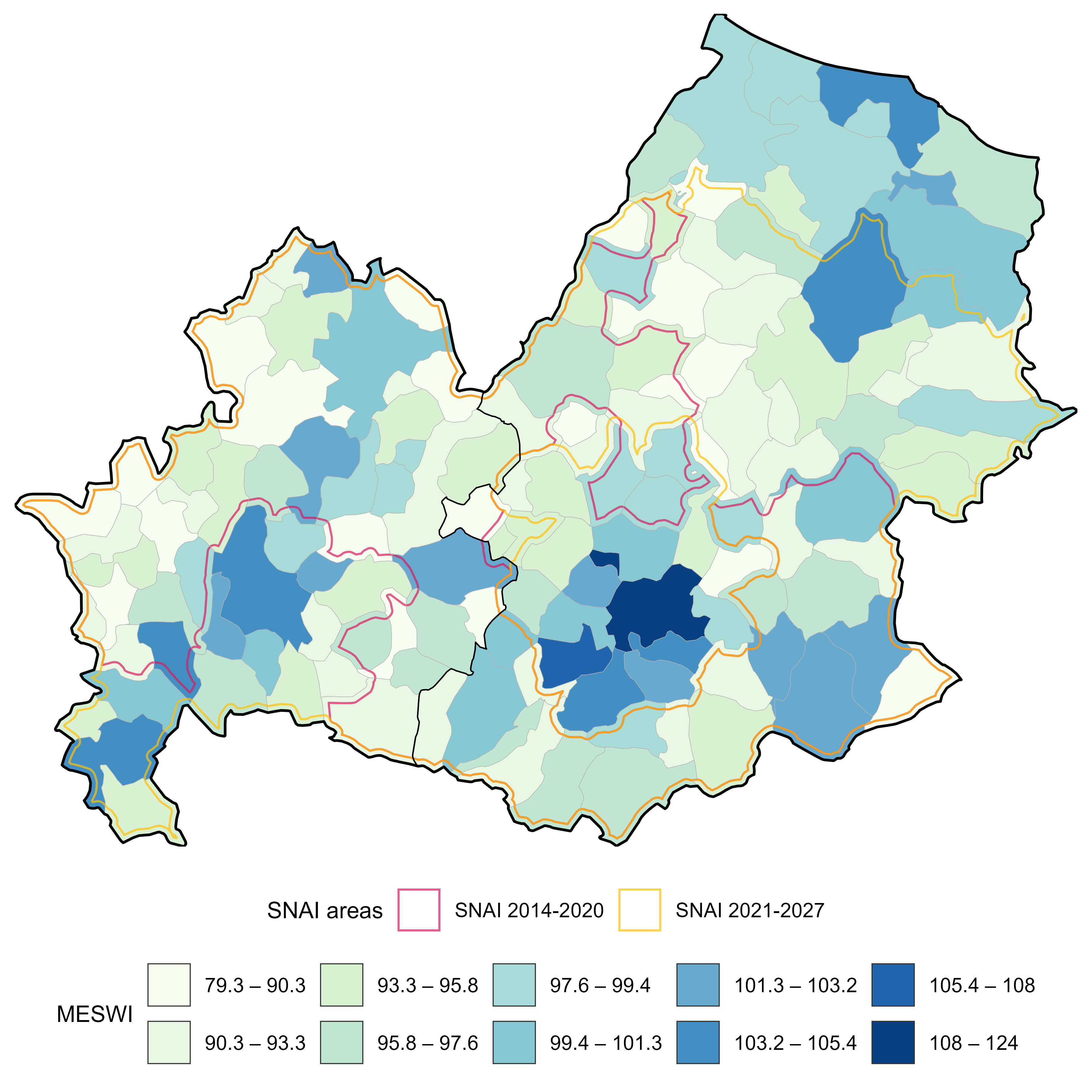} &
\includegraphics[width=\linewidth]{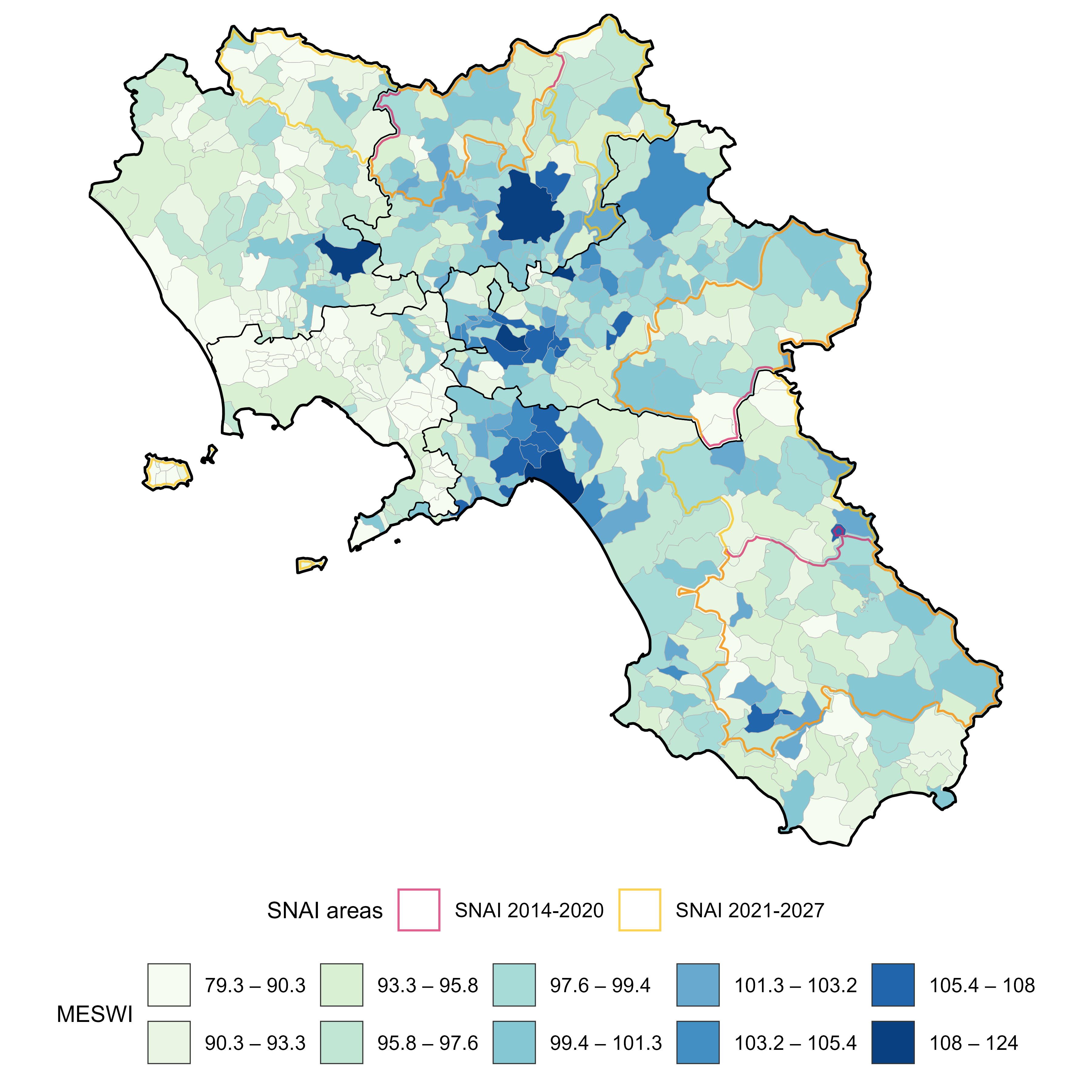}
\\[2mm]

{\small\textbf{Apulia}} &
{\small\textbf{Basilicata}} &
{\small\textbf{Calabria}}
\\[-1mm]

\includegraphics[width=\linewidth]{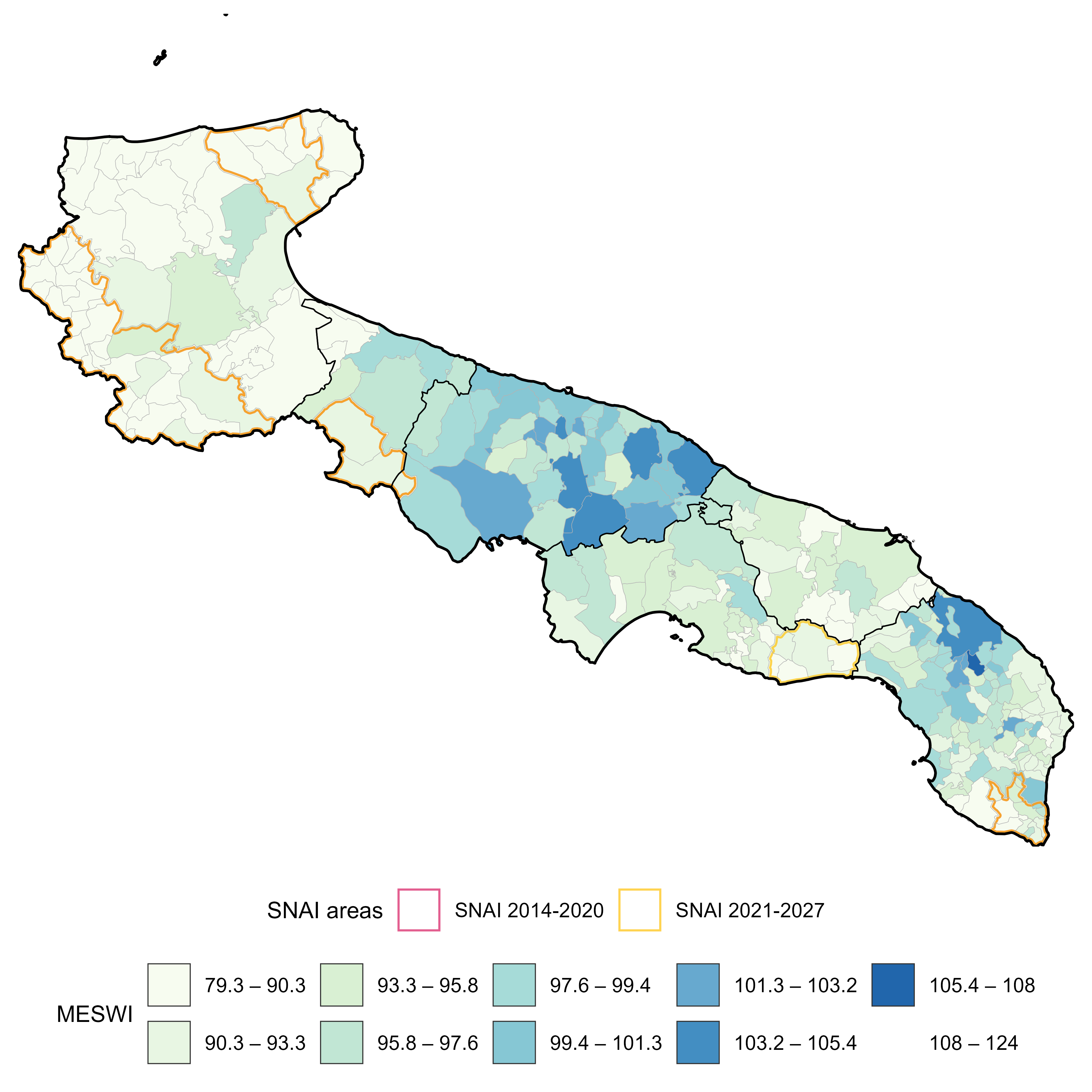} &
\includegraphics[width=\linewidth]{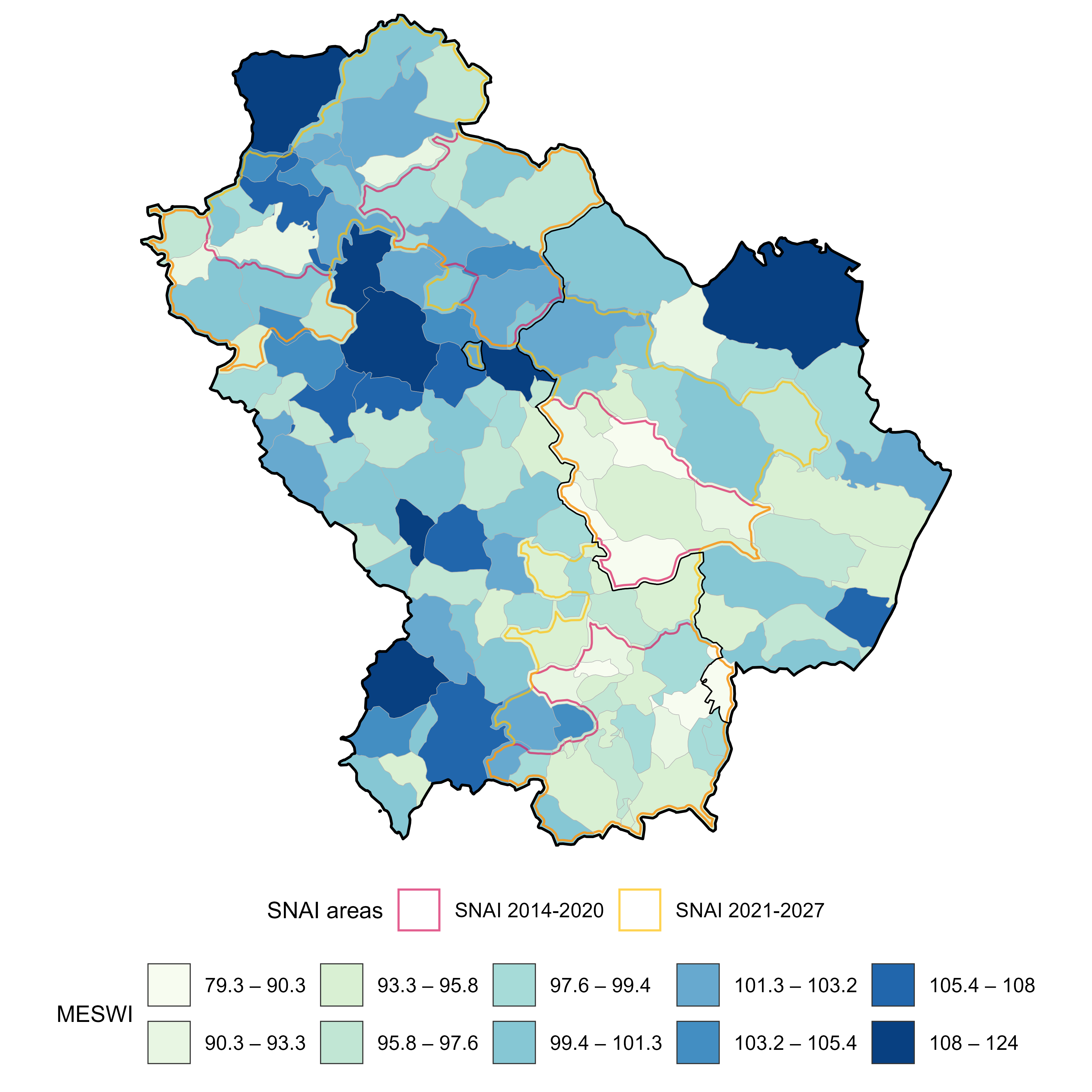} &
\includegraphics[width=\linewidth]{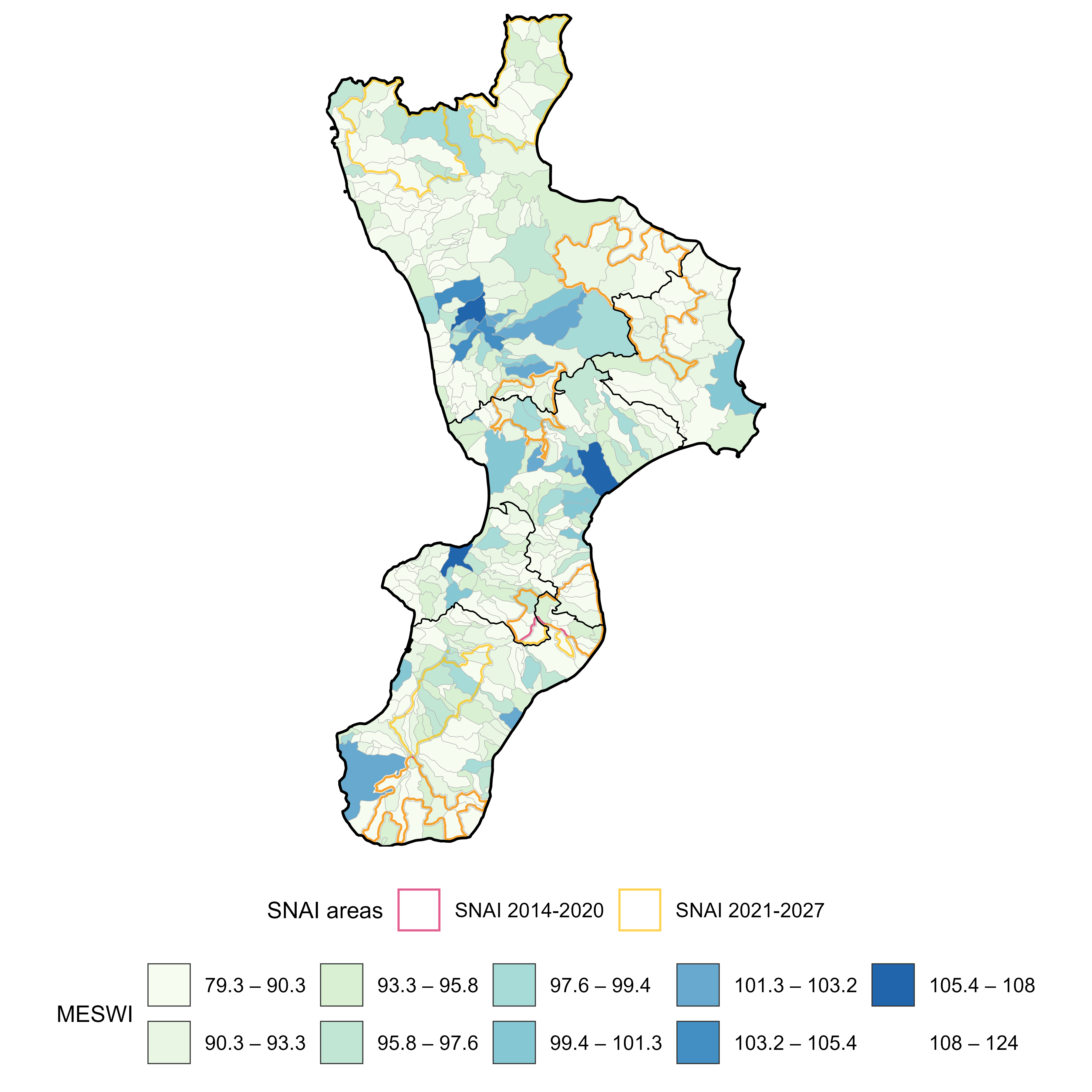}

\end{tabular}
\end{figure}


\begin{figure}[H]
\centering
\ContinuedFloat
{\small\textit{(continued)}}

\begin{tabular}{
    >{\centering\arraybackslash}p{0.32\textwidth}
    >{\centering\arraybackslash}p{0.32\textwidth}
    >{\centering\arraybackslash}p{0.32\textwidth}
}

{\small\textbf{Sicily}} &
{\small\textbf{Sardinia}} &
{}
\\[-1mm]

\includegraphics[width=\linewidth]{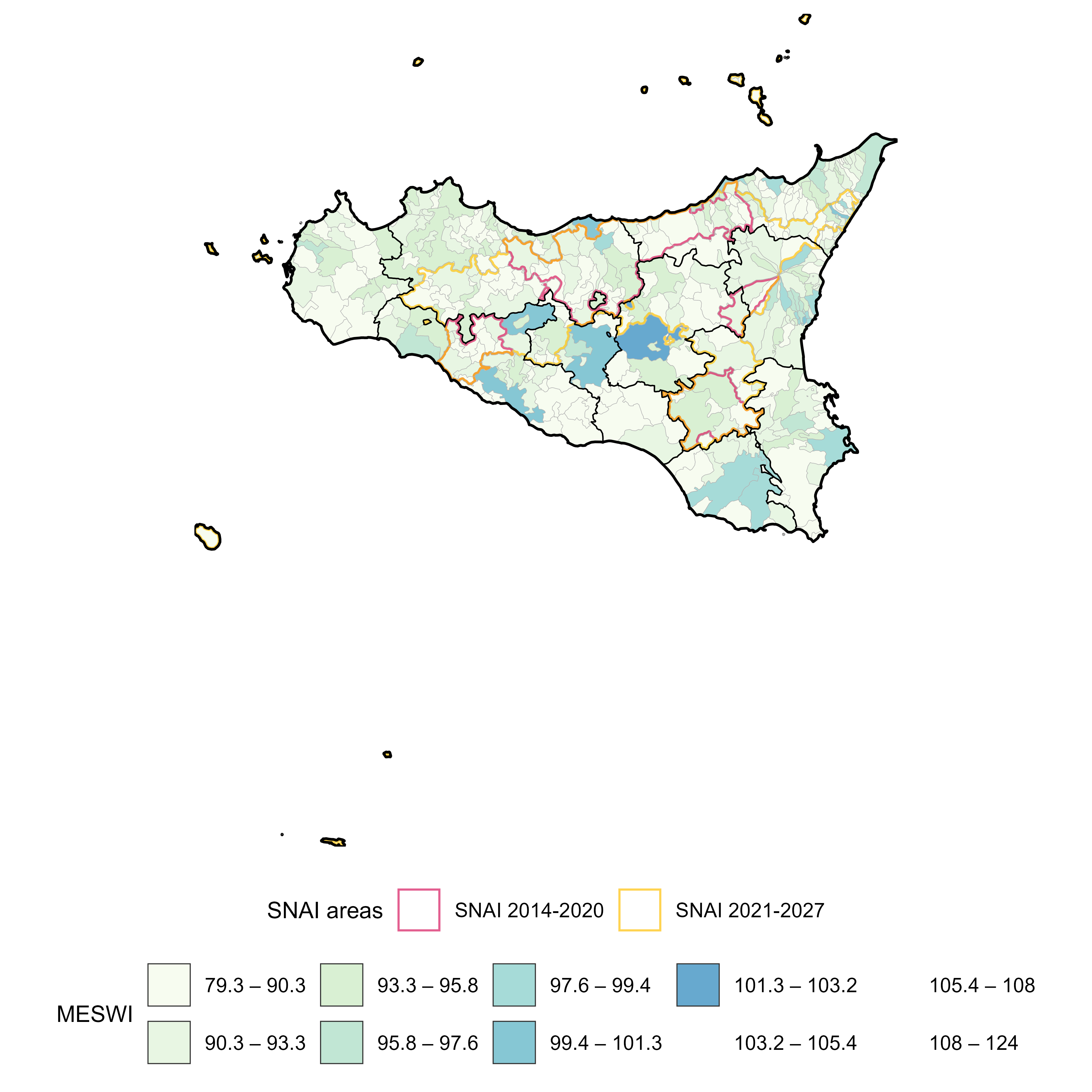} &
\includegraphics[width=\linewidth]{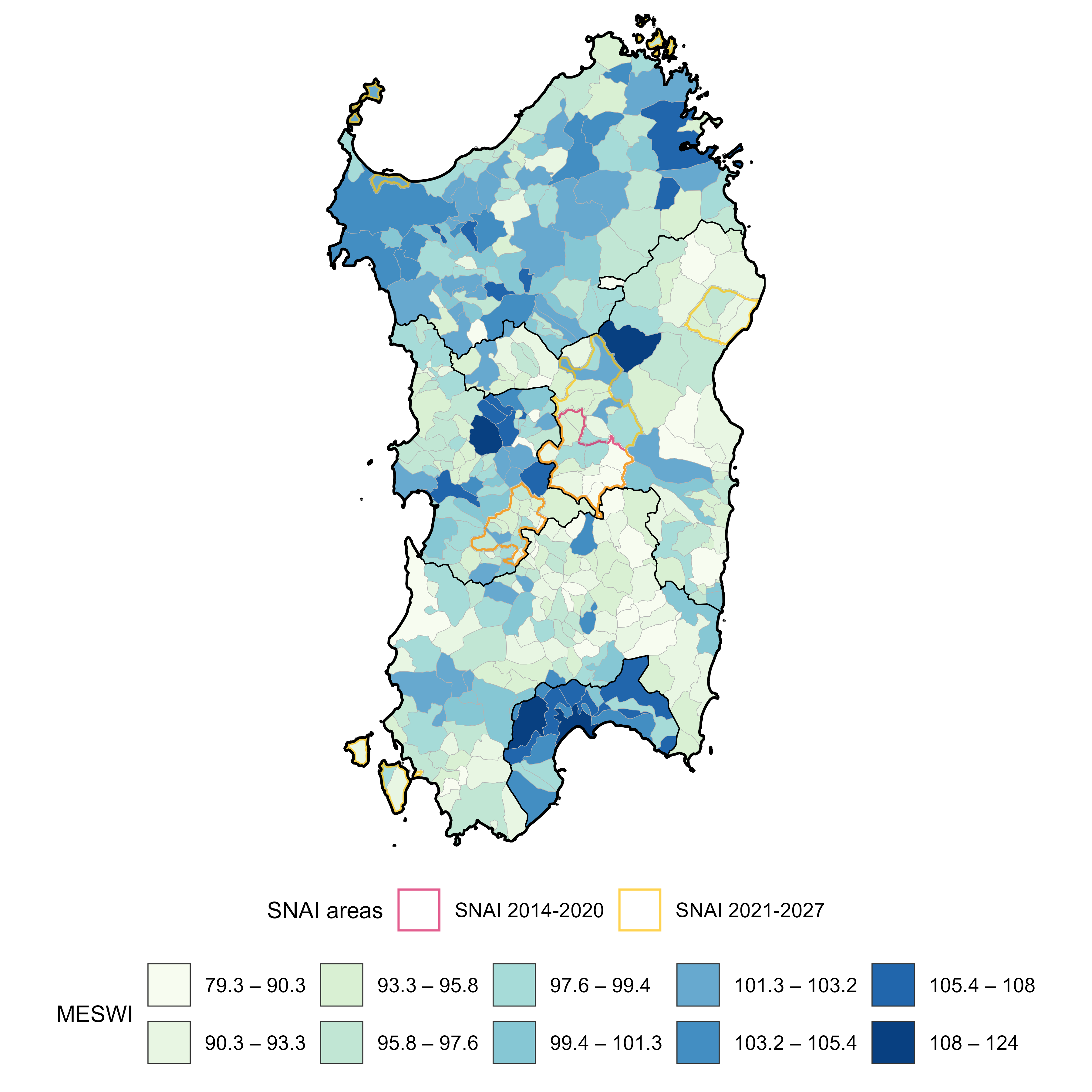} &
{}

\end{tabular}
\end{figure}

\begin{figure}[H]
\centering
\caption{\textit{Distribution of MESWI by macro-region and inner-area status (after excluding safety and subjective well-being dimensions)}}
\includegraphics[width=0.94\textwidth]{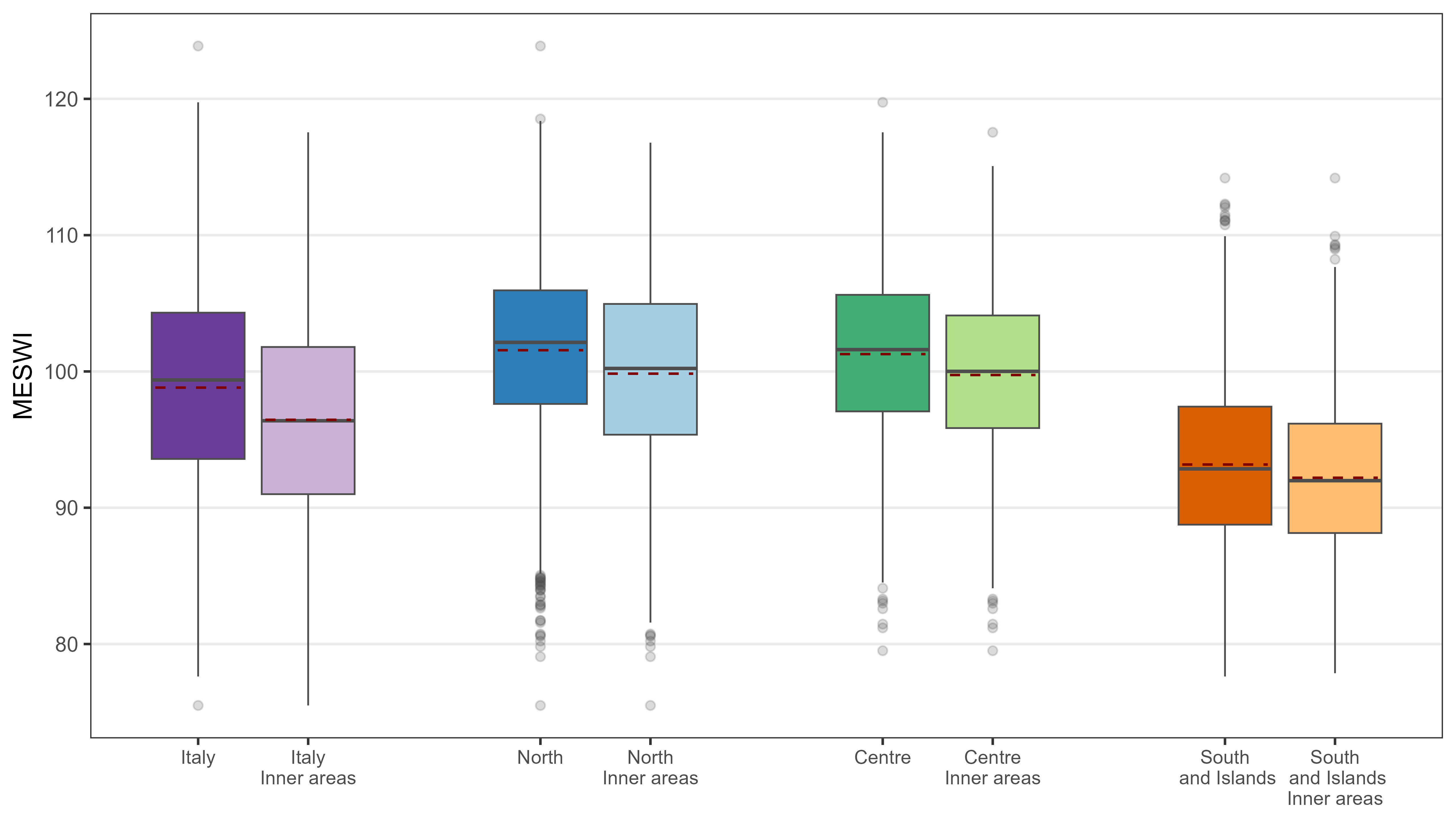}
\label{fig:robust-box}
\begin{minipage}{0.92\textwidth}
\tiny
Notes: inner areas comprise intermediate, peripheral and ultra-peripheral municipalities, together with municipalities belonging to SNAI project areas under the 2021--2027 classification. The solid grey line inside each box represents the median, while the dashed dark-red line indicates the mean. Boxes report the interquartile range, whiskers extend to 1.5 times the interquartile range, and points denote outlying observations.
\end{minipage}
\end{figure}

\begin{figure}[H]
\centering
\caption{\textit{Mean domain profiles by SNAI class (after excluding safety and subjective well-being dimensions)}}
\includegraphics[width=0.74\textwidth]{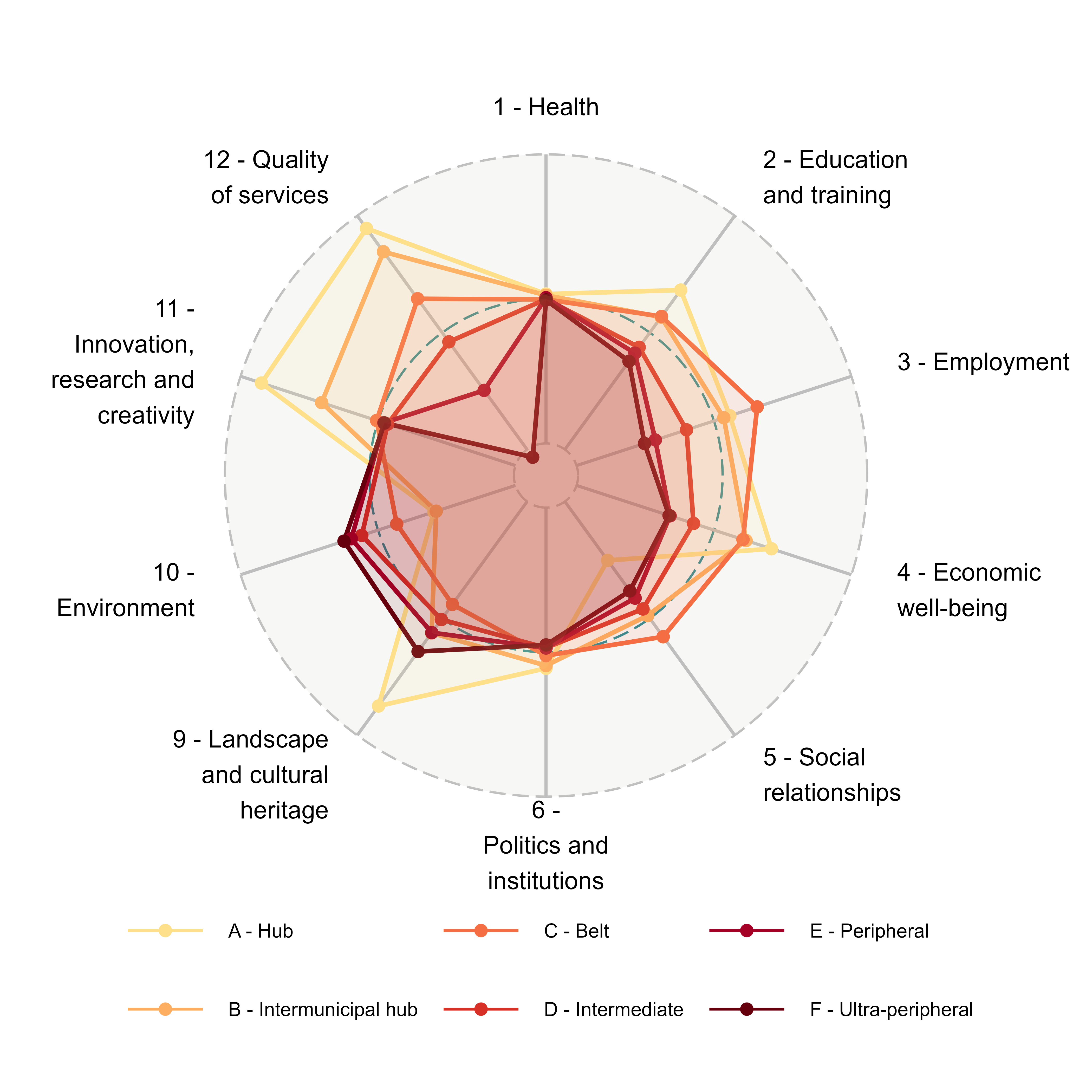}
\label{fig:robust-radar}
\end{figure}

\begin{figure}[H]
\centering
\caption{\textit{MESWI deciles and SNAI project areas (after excluding safety and subjective well-being dimensions)}}
\includegraphics[width=0.99\textwidth]{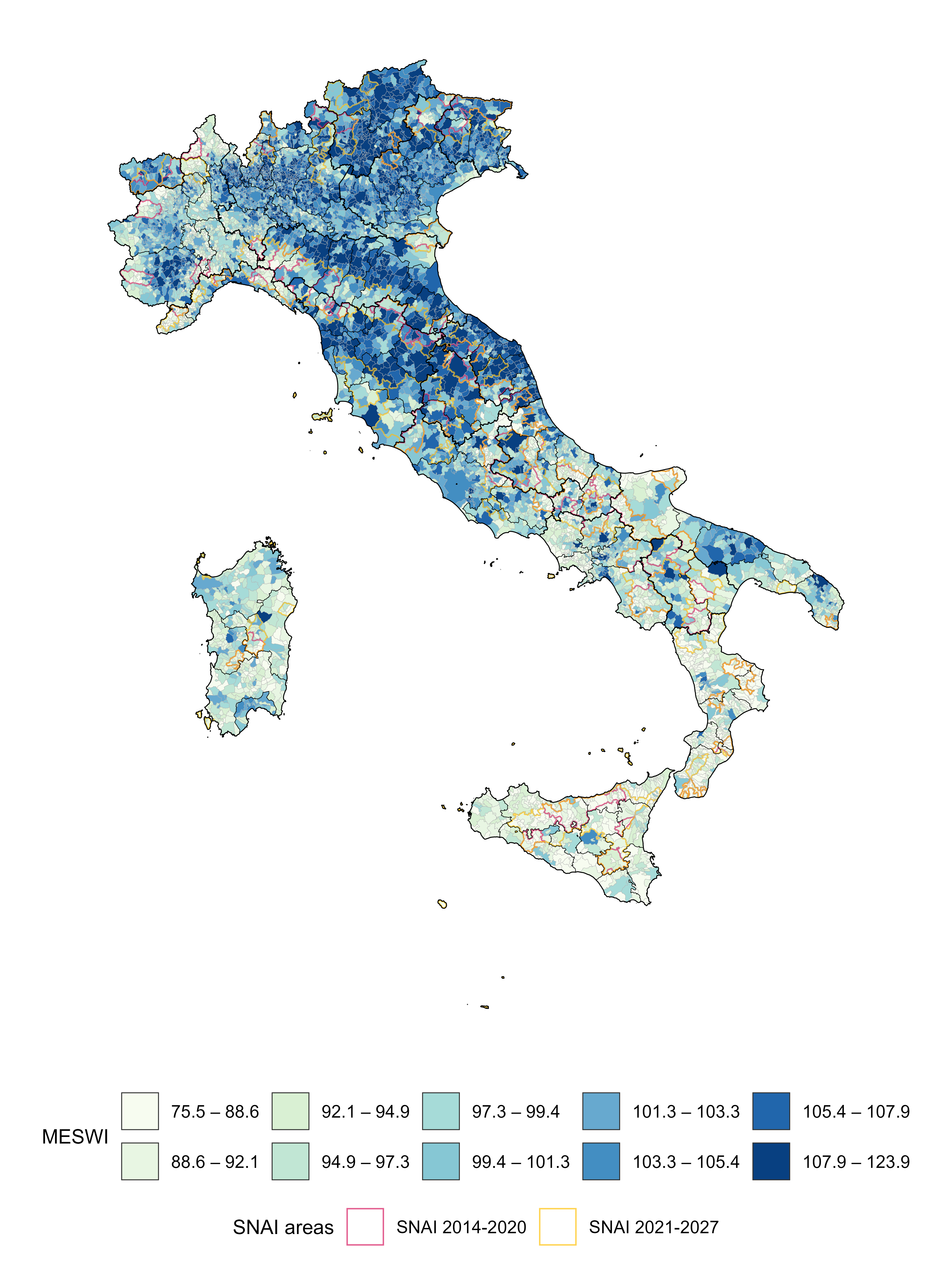}
\label{fig:robust-map}
\begin{minipage}{0.90\textwidth}
\tiny
Notes: municipal fill colours represent national MESWI deciles. Pink and yellow outlines identify SNAI project areas in the 2014--2020 and 2021--2027 cycles, respectively.\end{minipage}
\end{figure}

\begin{landscape}
\scriptsize
\setlength{\LTleft}{0pt}
\setlength{\LTright}{0pt}
\begin{longtable}{@{}p{2.5cm}p{5cm}c p{2.4cm}rrrrrrr c@{}}
\caption{\textit{Elementary indicators, sources, polarity and descriptive statistics}}\label{tab:indicator-details}\\
\toprule
\multicolumn{1}{c}{\textbf{Domain}} &
\multicolumn{1}{c}{\textbf{Indicator}} &
\multicolumn{1}{c}{\textbf{Year}} &
\multicolumn{1}{c}{\textbf{Source}} &
\multicolumn{1}{c}{\textbf{Mean}} &
\multicolumn{1}{c}{\textbf{SD}} &
\multicolumn{1}{c}{\textbf{Min}} &
\multicolumn{1}{c}{\textbf{Q1}} &
\multicolumn{1}{c}{\textbf{Median}} &
\multicolumn{1}{c}{\textbf{Q3}} &
\multicolumn{1}{c}{\textbf{Max}} &
\multicolumn{1}{c}{\textbf{Sign}} \\
\midrule
\endfirsthead
\multicolumn{12}{c}{\tablename\ \thetable{} -- continued}\\
\toprule
\multicolumn{1}{c}{\textbf{Domain}} &
\multicolumn{1}{c}{\textbf{Indicator}} &
\multicolumn{1}{c}{\textbf{Year}} &
\multicolumn{1}{c}{\textbf{Source}} &
\multicolumn{1}{c}{\textbf{Mean}} &
\multicolumn{1}{c}{\textbf{SD}} &
\multicolumn{1}{c}{\textbf{Min}} &
\multicolumn{1}{c}{\textbf{Q1}} &
\multicolumn{1}{c}{\textbf{Median}} &
\multicolumn{1}{c}{\textbf{Q3}} &
\multicolumn{1}{c}{\textbf{Max}} &
\multicolumn{1}{c}{\textbf{Sign}} \\
\midrule
\endhead
\midrule\multicolumn{12}{r}{Continued on next page}\endfoot
\bottomrule\endlastfoot
Health & Mean age & 2024 & \cite{istat_amdc} & 48.4 & 3.2 & 37.6 & 46.3 & 48.1 & 50.2 & 65.2 & + \\
Health & Mortality rate (per thousand) & 2024 & \cite{istat_amdc} & 12.9 & 5.6 & 0.0 & 9.5 & 11.9 & 14.9 & 82.5 & - \\
Health & Birth rate (per thousand) & 2024 & \cite{istat_amdc} & 5.7 & 2.7 & 0.0 & 4.3 & 5.7 & 7.1 & 42.1 & + \\
Education and training & Young people not in employment, education or training (NEET), \% [municipalities above 5,000 residents] & 2022 & \cite{istat_amdc} & 18.1 & 4.2 & 7.8 & 14.8 & 17.2 & 21.0 & 52.1 & - \\
Education and training & INVALSI literacy score [provincial capitals] & 2024 & \cite{istat_amdc} & 197.5 & 6.1 & 179.3 & 192.9 & 198.3 & 202.3 & 211.0 & + \\
Education and training & INVALSI mathematics score [provincial capitals] & 2024 & \cite{istat_amdc} & 198.5 & 9.3 & 173.9 & 192.3 & 199.2 & 205.3 & 217.9 & + \\
Education and training & Population with upper-secondary education, \% & 2023 & \cite{istat_amdc} & 74.5 & 7.3 & 40.4 & 70.2 & 75.1 & 79.5 & 100.0 & + \\
Education and training & Population with tertiary education, \% & 2023 & \cite{istat_amdc} & 23.3 & 6.0 & 0.0 & 19.4 & 22.9 & 26.9 & 52.2 & + \\
Work and work-life balance & Workers in unstable employment, \% & 2022 & \cite{istat_amdc} & 12.6 & 3.8 & 5.7 & 9.8 & 11.4 & 15.0 & 36.5 & - \\
Work and work-life balance & Unemployment rate, \% & 2023 & \cite{istat_amdc} & 7.4 & 3.5 & 0.0 & 4.7 & 6.3 & 9.8 & 39.8 & - \\
Work and work-life balance & Inactivity rate, \% & 2023 & \cite{istat_amdc} & 50.2 & 6.1 & 28.8 & 45.7 & 49.8 & 54.4 & 75.9 & - \\
Work and work-life balance & Employment rate, \% & 2023 & \cite{istat_amdc} & 46.3 & 6.9 & 22.7 & 41.2 & 46.9 & 51.5 & 70.5 & + \\
Economic well-being & Personal income taxpayers with total income below EUR 10,000, \% & 2023 & \cite{istat_amdc} & 25.7 & 8.2 & 11.9 & 18.9 & 23.5 & 31.6 & 69.3 & - \\
Economic well-being & Households with low work intensity, \% [municipalities above 5,000 residents] & 2022 & \cite{istat_amdc} & 44.2 & 5.7 & 24.4 & 40.4 & 43.8 & 47.7 & 72.8 & - \\
Economic well-being & Single-earner households with children aged under six, \% [municipalities above 5,000 residents] & 2022 & \cite{istat_amdc} & 20.5 & 4.4 & 8.6 & 17.5 & 20.6 & 23.2 & 45.8 & - \\
Economic well-being & Mean taxable income (EUR) & 2023 & \cite{istat_amdc} & 21247.6 & 4176.7 & 7318.8 & 18168.4 & 21414.2 & 23978.0 & 94505.4 & + \\
Social relationships & Average household size & 2023 & \cite{istat_amdc} & 2.2 & 0.2 & 1.1 & 2.1 & 2.2 & 2.3 & 2.9 & + \\
Social relationships & Households with children, \% [municipalities above 5,000 residents] & 2022 & \cite{istat_amdc} & 24.5 & 4.4 & 5.0 & 22.0 & 24.5 & 27.0 & 46.4 & + \\
Social relationships & Households with minors, \% [municipalities above 5,000 residents] & 2022 & \cite{istat_amdc} & 18.9 & 2.5 & 10.3 & 17.6 & 18.8 & 20.3 & 37.4 & + \\
Politics and institutions & Voter turnout, \% & 2024 & \cite{istat_amdc} & 62.6 & 10.3 & 11.9 & 56.8 & 63.2 & 69.2 & 91.9 & + \\
Politics and institutions & Mean age of municipal office-holders & 2024 & \cite{istat_amdc} & 51.3 & 7.2 & 27.3 & 46.5 & 51.0 & 55.8 & 83.0 & - \\
Politics and institutions & Municipal Administration Quality Index: bureaucracy & 2022 & \cite{cerqua2025municipal} & 99.4 & 5.3 & 68.5 & 97.7 & 100.3 & 102.4 & 118.2 & + \\
Politics and institutions & Municipal Administration Quality Index: economy & 2022 & \cite{cerqua2025municipal} & 103.7 & 3.9 & 81.9 & 102.0 & 104.2 & 106.2 & 118.9 & + \\
Politics and institutions & Municipal Administration Quality Index: politics & 2022 & \cite{cerqua2025municipal} & 107.6 & 9.4 & 83.7 & 102.9 & 107.2 & 114.2 & 135.6 & + \\
Safety & Other reported lethal offences per 100,000 residents [province] & 2023 & \cite{ISTAT2025a} & 2.9 & 1.2 & 1.0 & 2.0 & 3.0 & 3.0 & 8.0 & - \\
Safety & Reported pickpocketing offences per 100,000 residents [province] & 2023 & \cite{ISTAT2025a} & 139.1 & 178.4 & 3.0 & 41.0 & 88.0 & 162.0 & 906.0 & - \\
Safety & Reported residential burglaries per 100,000 residents [province] & 2023 & \cite{ISTAT2025a} & 240.8 & 95.3 & 55.0 & 166.0 & 263.0 & 314.0 & 482.0 & - \\
Safety & Reported robberies per 100,000 residents [province] & 2023 & \cite{ISTAT2025a} & 32.7 & 25.8 & 3.0 & 15.0 & 28.0 & 40.0 & 137.0 & - \\
Safety & Road mortality outside urban areas, \% [province] & 2023 & \cite{ISTAT2025a} & 4.6 & 2.1 & 0.0 & 3.0 & 4.0 & 6.0 & 11.0 & - \\
Safety & Intentional homicides per 1,000 residents [province] & 2024 & \cite{ISTAT2025a} & 0.7 & 0.6 & 0.0 & 0.0 & 1.0 & 1.0 & 3.0 & - \\
Subjective well-being & Negative assessment of future prospects, \% [region] & 2024 & \cite{ISTAT2025b} & 12.3 & 2.0 & 8.3 & 10.9 & 12.0 & 13.5 & 16.5 & - \\
Subjective well-being & Positive assessment of future prospects, \% [region] & 2024 & \cite{ISTAT2025b} & 30.8 & 2.4 & 25.7 & 28.6 & 30.8 & 33.0 & 34.7 & + \\
Subjective well-being & Satisfaction with leisure time, \% [region] & 2024 & \cite{ISTAT2025b} & 67.2 & 3.1 & 60.4 & 65.8 & 67.8 & 69.6 & 80.6 & + \\
Subjective well-being & Life satisfaction, \% [region] & 2024 & \cite{ISTAT2025b} & 47.2 & 4.2 & 37.6 & 46.7 & 47.5 & 48.2 & 69.4 & + \\
Landscape and cultural heritage & Libraries registered in the national library register per 100,000 residents & 2024 & \cite{istat_amdc} & 50.7 & 95.1 & 0.0 & 5.2 & 23.0 & 58.2 & 2061.9 & + \\
Landscape and cultural heritage & Museums, galleries, archaeological sites and monuments per 100,000 residents & 2022 & \cite{istat_amdc} & 19.3 & 80.0 & 0.0 & 0.0 & 0.0 & 5.1 & 1851.9 & + \\
Landscape and cultural heritage & Visitors to museums, galleries, archaeological sites and monuments per 100,000 residents & 2022 & \cite{istat_amdc} & 155.7 & 5425.7 & 0.0 & 0.0 & 0.0 & 2.6 & 477777.8 & + \\
Environment & Land take, \% & 2024 & \cite{istat_amdc} & 10.1 & 10.3 & 0.3 & 3.4 & 6.6 & 12.7 & 91.8 & - \\
Environment & Tree cover, \% & 2020 & \cite{ESA2026} & 46.1 & 29.0 & 0.0 & 17.8 & 47.6 & 71.7 & 99.5 & + \\
Environment & Noise pollution [provincial capitals] & 2022 & \cite{istat_amdc} & 2.5 & 3.8 & 0.0 & 0.0 & 1.2 & 3.0 & 19.9 & - \\
Environment & Urban air pollution (PM10) [provincial capitals] & 2023 & \cite{istat_amdc} & 23.8 & 19.0 & 0.0 & 9.0 & 20.0 & 33.0 & 70.0 & - \\
Environment & Separate collection of municipal waste, \% & 2024 & \cite{istat_amdc} & 70.2 & 15.0 & 0.0 & 63.9 & 73.1 & 80.7 & 100.0 & + \\
Innovation, research and creativity & Fibre-to-the-home (FTTH) network coverage, \% & 2024 & \cite{istat_amdc} & 31.8 & 23.1 & 0.0 & 10.0 & 33.1 & 49.0 & 95.9 & + \\
Innovation, research and creativity & Productive specialisation in high-technology sectors, \% & 2023 & \cite{istat_amdc} & 2.1 & 3.8 & 0.0 & 0.8 & 1.2 & 1.9 & 72.3 & + \\
Quality of services & Children enrolled in municipal childcare services, \% & 2023 & \cite{istat_amdc} & 15.3 & 27.3 & 0.0 & 0.0 & 7.5 & 23.8 & 1133.3 & + \\
Quality of services & Travel time to the nearest hospital (minutes) & 2025 & \cite{InstantAnalytics2025} & 30.1 & 16.9 & 1.2 & 18.4 & 26.4 & 38.3 & 335.3 & - \\
Quality of services & Travel time to the nearest upper-secondary school (minutes) & 2025 & \cite{InstantAnalytics2025} & 18.1 & 9.7 & 0.5 & 12.0 & 16.2 & 22.1 & 163.7 & - \\
Quality of services & Travel time to the nearest railway station (minutes) & 2025 & \cite{InstantAnalytics2025} & 26.6 & 17.4 & 0.4 & 14.6 & 22.2 & 34.5 & 331.5 & - \\
Quality of services & Municipal expenditure on social interventions and services per resident (EUR) & 2022 & \cite{istat_amdc} & 119.7 & 112.4 & 0.0 & 56.3 & 92.1 & 139.8 & 1462.3 & + \\
\end{longtable}
\begin{minipage}{\linewidth}
\tiny
\vspace{-.2cm}
Notes: NEET = not in employment, education or training; FTTH = fibre to the home. A plus sign indicates that higher values are associated with higher well-being; a minus sign indicates the opposite. Statistics refer to the municipal analysis file after the spatial imputations described in the text and before polarity reversal, logarithmic transformation and standardisation.
\end{minipage}
\end{landscape}

\scriptsize
\begin{longtable}{@{}l r rrr rrr@{}}
\caption{\textit{Mean MESWI by region and SNAI status}}
\label{tab:regions}\\
\toprule
& & \multicolumn{3}{c}{\textbf{SNAI 2014--2020}} & \multicolumn{3}{c}{\textbf{SNAI 2021--2027}}\\
\cmidrule(lr){3-5}\cmidrule(lr){6-8}
\multicolumn{1}{c}{\textbf{Area}} & Total & Project areas & Inner areas & Non-inner & Project areas & Inner areas & Non-inner \\
\midrule
\endfirsthead
\multicolumn{8}{c}{\tablename\ \thetable{} -- continued}\\
\toprule
& & \multicolumn{3}{c}{SNAI 2014--2020} & \multicolumn{3}{c}{SNAI 2021--2027}\\
\cmidrule(lr){3-5}\cmidrule(lr){6-8}
Area & Total & Project areas & Inner areas & Non-inner & Project areas & Inner areas & Non-inner \\
\midrule
\endhead
\midrule\multicolumn{8}{r}{Continued on next page}\endfoot
\bottomrule\endlastfoot
Abruzzo & 96.41 & 93.18 & 95.06 & 100.43 & 94.27 & 94.60 & 100.22 \\
Basilicata & 98.94 & 95.71 & 98.51 & 109.74 & 97.14 & 98.42 & 104.91 \\
Calabria & 91.65 & 90.05 & 90.86 & 94.97 & 90.48 & 90.41 & 94.44 \\
Campania & 95.71 & 96.46 & 95.58 & 95.76 & 95.41 & 95.56 & 95.73 \\
Emilia-Romagna & 99.36 & 94.48 & 97.04 & 100.96 & 95.83 & 97.33 & 101.41 \\
Friuli--Venezia Giulia & 104.58 & 103.56 & 103.44 & 105.32 & 103.35 & 103.72 & 104.99 \\
Lazio & 96.57 & 93.76 & 96.17 & 98.17 & 94.90 & 95.73 & 97.88 \\
Liguria & 97.02 & 95.84 & 95.15 & 98.28 & 95.19 & 94.90 & 99.28 \\
Lombardy & 104.52 & 101.82 & 102.95 & 105.27 & 102.32 & 103.04 & 105.27 \\
Marche & 101.87 & 100.30 & 99.87 & 103.61 & 100.22 & 100.86 & 102.83 \\
Molise & 95.36 & 94.41 & 94.55 & 99.39 & 94.45 & 94.71 & 98.91 \\
Piemonte & 97.33 & 93.24 & 95.41 & 98.51 & 93.31 & 95.17 & 98.40 \\
Apulia & 93.02 & 88.19 & 90.86 & 95.56 & 88.18 & 90.85 & 96.04 \\
Sardegna & 96.98 & 94.64 & 96.30 & 100.64 & 95.15 & 96.22 & 98.81 \\
Sicily & 90.91 & 90.02 & 90.27 & 92.79 & 89.95 & 90.60 & 92.11 \\
Tuscany & 101.66 & 97.45 & 99.58 & 103.44 & 99.31 & 100.13 & 104.09 \\
Trentino-Alto Adige & 109.83 & 109.37 & 109.20 & 111.59 & 108.05 & 108.98 & 112.71 \\
Umbria & 103.24 & 101.09 & 102.24 & 105.45 & 102.10 & 102.28 & 105.52 \\
Valle d'Aosta & 106.73 & 106.32 & 106.29 & 107.05 & 106.33 & 105.78 & 108.53 \\
Veneto & 103.59 & 101.37 & 102.45 & 104.18 & 100.98 & 101.50 & 104.09 \\
\textbf{North} & 102.06 & 99.11 & 100.86 & 102.83 & 100.00 & 100.81 & 102.80 \\
\textbf{Centre} & 99.87 & 97.14 & 98.15 & 102.51 & 98.40 & 98.68 & 101.61 \\
\textbf{South and Islands} & 94.48 & 93.00 & 93.71 & 96.29 & 93.07 & 93.56 & 96.25 \\
\textbf{Italy} & 99.34 & 95.77 & 97.29 & 101.54 & 96.58 & 97.27 & 101.38 \\
\end{longtable}
\begin{minipage}{\linewidth}
\tiny
Notes: entries are unweighted means across municipalities. Inner areas comprise municipalities classified as intermediate, peripheral or ultra-peripheral, together with municipalities included in SNAI project areas.
\end{minipage}


\end{document}